\documentclass[reprint, aps, prb, amsmath, showpacs, superscriptaddress]{revtex4-2}
\usepackage{graphicx}
\usepackage{epsfig}
\usepackage[cp1251]{inputenc}
\usepackage[english]{babel}
\usepackage{amsmath}
\usepackage{amssymb}
\usepackage{amstext}
\usepackage{color}
\usepackage{float}
\usepackage{epstopdf}
\usepackage{hyperref}
\begin{document}

\title{Transition magnon modes in thin ferromagnetic nanogratings}

\author{S. M. Kukhtaruk}
\email{serhii.kukhtaruk@tu-dortmund.de}
\affiliation{Experimentelle Physik 2, Technische Universit\"{a}t Dortmund, Otto-Hahn-Str. 4a, 44227 Dortmund, Germany.}
\affiliation{Department of Theoretical Physics, V.E. Lashkaryov Institute of Semiconductor Physics, 03028 Kyiv, Ukraine. }
\author{A. W. Rushforth}
\affiliation{School of Physics and Astronomy, University of Nottingham, Nottingham NG7 2RD, United Kingdom.}
\author{F.~Godejohann}
\affiliation{Experimentelle Physik 2, Technische Universit\"{a}t Dortmund, Otto-Hahn-Str. 4a, 44227 Dortmund, Germany.}
\author{A. V. Scherbakov}
\affiliation{Experimentelle Physik 2, Technische Universit\"{a}t Dortmund, Otto-Hahn-Str. 4a, 44227 Dortmund, Germany.}
\author{M. Bayer}
\affiliation{Experimentelle Physik 2, Technische Universit\"{a}t Dortmund, Otto-Hahn-Str. 4a, 44227 Dortmund, Germany.}

\begin{abstract}
This work presents micromagnetic simulations in ferromagnetic nanogratings for the full range of directions of an applied in-plane external magnetic field. We focus on the modification of the magnon mode characteristics when the magnetic field orientation is gradually changed between the classical Damon-Eshbach (DE) and backward-volume (BV) geometries. We found that in a specific range of field directions, the magnon mode parameters differ significantly from the parameters in the classical cases, namely, the modes are characterized by complex spatial distributions and have low group velocities. The center of this range corresponds to the direction of the external magnetic field, which gives the maximal nonuniform distribution of the static magnetization in the nanogratings.
\end{abstract}

\maketitle

\section{INTRODUCTION}

Magnonics is a rapidly developing direction of modern magnetism that uses spin waves (or magnons) for data transfer and processing~\cite{Lenk_2011, Krawczyk_2014, Chumak_2015}. Structures with spatially periodic magnetic properties, referred to as magnonic crystals~\cite{Nikitov_2001,Kostylev_2004,Guslienko_2005,Gubbiotti_2005,Gubbiotti_2007,Kostylev_2008,Polushkin_2008,Mruczkiewicz_Comsol,Mruczkiewicz_2013,Lisenkov_2015,Rychly_2015}, and periodically patterned ferromagnetic films, referred to as surface modulated magnonic crystals~\cite{Chumak_2008,Chumak_2009,Chumak_2009APL,Serga_2010,Landeros_2012,Kakazei_2014,Aranda_2014,Bessonov_2015,Langer_2017,Gallardo_2018,Langer_2019,Gallardo_2019}, take a central role in this field. The simplest pattern is a periodic set of grooves and wires that form a\mbox{ (nano-) grating} (NG) structure which, despite its simple design, already shows prospective potential for multiple applications, e.g. magnonic transistors~\cite{Chumak_2014}, switches~\cite{Khitun_2010,Vogt_2014,Balinskiy_2018}, filters~\cite{Chumak_2008,Kim_2009}, grating couplers~\cite{Yu_2013}, magnetic field generators~\cite{Salasyuk_2018} and detectors~\cite{Inoue_2011}.

An essential feature of NG structures is the rich spectrum of magnon eigenmodes, which can be tuned by an external magnetic field, $\textstyle{{\bf H}_{\rm ext}}$. It consists of the modes with wave vector along the reciprocal wave vector of the NG (perpendicular to the grooves) with energy dispersions and spatial distributions significantly different from the case of a plain film (see e.g.~\cite{Lisenkov_2015,Langer_2019}). The dispersions determine the magnon modes' group velocities as well as the magnon band gaps' spectral positions and widths. The spatial profiles of the magnon modes define their excitation selectivity and the interaction with other periodic excitations, such as electromagnetic~\cite{Kostylev_2004,Gubbiotti_2005,Gubbiotti_2007,Kostylev_2008,Kakazei_2014,Aranda_2014,Gallardo_2018,Langer_2019} or elastic~\cite{Bombeck_2012,Verba_2018,Godejohann_2020} waves. The excitation and interaction efficiencies depend on the spatial matching of the magnon modes and other wave excitations, and can be entirely suppressed in case of poor matching.

An essential fundamental phenomenon in magnonics is \textit{nonreciprocity}, i.e. the difference in localization~\cite{DE_1961} and/or amplitude~\cite{Schneider_2008,Kostylev_2013} and/or frequency~\cite{Lisenkov_2015,Rychly_2015,Bessonov_2015,Mruczkiewicz_2017,Gallardo_2019} of the spin waves with opposite directions of their wave vectors. In NG structures, magnons can possess asymmetric dispersions with indirect band gaps in the Brillouin zone~\cite{Lisenkov_2015,Bessonov_2015}. In this case, spin waves with Bloch wavevectors at the center of the Brillouin zone are propagating waves. Thus, nonreciprocity provides the ability to excite propagating spin waves employing spatially homogeneous \mbox{excitations.}

The main experimental tools for studying magnons in NG structures are Brillouin light scattering~\cite{Kostylev_2004,Gubbiotti_2005,Gubbiotti_2007,Kostylev_2008,Krawczyk_2014,Gubbiotti_2021} and ferromagnetic resonance (FMR)~\cite{Farle_1998,Landeros_2012,Kakazei_2014,Aranda_2014,Gallardo_2018,Langer_2019} spectroscopy. Traditionally, only two directions of $\textstyle{{\bf H}_{\rm ext}}$ are studied: either perpendicular (the Damon-Eshbach (DE) geometry) or parallel (the backward-volume (BV) geometry) to the magnon wavevector. However, to the best of our knowledge, so far there is no information available about the main magnon characteristics in NG structures for intermediate directions of $\textstyle{{\bf H}_{\rm ext}}$.

While magnons in plain ferromagnetic films are well studied for any magnetic field direction~\cite{DE_1961, Wames_1970}, the introduction of grooves not only fixes the magnon wave vector direction, but also leads to the generation of both static and dynamic demagnetizing fields, which strongly influence the magnon characteristics. For instance, the magnon spatial profiles and dispersions at $\textstyle{45^{\circ}}$, i.e. in between the DE and BV geometries, are a priori not known, i.e., the question is whether this case is DE-like or BV-like? In addition, it is unclear how the transition between the DE and BV geometries is arranged, i.e., is the change of the mode characteristics smooth across the wide range or does it occur sharply for a specific magnetic field direction? Moreover, time-resolved experiments on excitation of a metallic ferromagnet by femtosecond laser pulses and detection of the coherent magnon response using the transient magneto-optical Kerr effect (MOKE)~\cite{Hiebert_1997,VanKampen_2002,Kats_2016,Salasyuk_2018,Scherbakov_2019,Khokhlov_2019} are typically performed at intermediate directions of $\textstyle{{\bf H}_{\rm ext}}$, where the excitation efficiency is maximal~\cite{Scherbakov_2019,Khokhlov_2019}. Thus, an analysis of the magnons at intermediate directions of $\textstyle{{\bf H}_{\rm ext}}$ providing answers to the above questions are required.

This paper presents a detailed theoretical analysis of magnons in a ferromagnetic NG for arbitrary directions of $\textstyle{{\bf H}_{\rm ext}}$. We focus on the main magnon characteristics including the spatial profiles, dispersions, and dependencies of the magnon frequencies on the direction and strength of $\textstyle{{\bf H}_{\rm ext}}$. We find that by changing the direction of the magnetic field continuously from DE geometry to BV geometry, the main magnon characteristics remain DE-like until the field direction enters a \textit{transition range}, where the magnon characteristics are neither DE-like nor BV-like. When exiting this transition range, the magnon characteristics can be considered to be BV-like. We show that both the static magnetization and the demagnetizing field are critically important for the transition. The position of the transition range is defined by the particular direction of the external magnetic field, which corresponds to the maximal nonuniform distribution of the static magnetization in the NG. The width of the transition range is $\textstyle{\sim25^{\circ}}$. Therefore, by changing the magnetic field direction around the transition range one can switch between three different types of magnon modes and choose the magnon modes with the desired characteristics.

For the calculations, we use the COMSOL Multiphysics$\textstyle{^\circledR}$ software~\cite{Comsol}. Our choice is based on the ability to solve and visualize the problem in both time and frequency domains with integration of a specific external impact, such as femtosecond laser excitation, monochromatic elastic waves, or picosecond strain pulses, which are widely used in time-resolved magnonic experiments (see e.g.~\cite{Hiebert_1997,VanKampen_2002,Kats_2016,Salasyuk_2018,Scherbakov_2019,Khokhlov_2019}).

The paper is organized as follows: The basic equations and used parameters are given in Sec.~\ref{Eqs}. The steady-state situation is considered in Sec~\ref{statics}. The magnon characteristics are discussed in Secs.~\ref{dynamics} and \ref{dispersion}. The validity and applications are discussed in Sec.~\ref{ValAndApp}. The conclusions are summarized in Sec.~\ref{conclusion}.

\section{BASIC EQUATIONS}\label{Eqs}
The system under study is a NG structure that consists of infinitely long parallel grooves of depth $\textstyle{h}$, separated by wires of width $\textstyle{w}$. The NG is produced in a ferromagnetic film of thickness $\textstyle{l}$, (see Fig.~1), located on a nonmagnetic substrate, which is not shown for simplicity. The NG period is $\textstyle{d}$, and the width of the grooves is $\textstyle{d-w}$. The external magnetic field, $\textstyle{{\bf H}_{\rm ext}}$, is applied in the structural plane of the ferromagnet at an angle $\textstyle{\varphi_{\rm H}}$ between the grooves and $\textstyle{{\bf H}_{\rm ext}}$.

The magnetization $\textstyle{{\bf M}}$ in the ferromagnetic NG structure is described by the Landau-Lifshitz-Gilbert equation (see, e.g.,~\cite{Landau_9,LLB_1984,Fesenko_2006}). It is convenient to introduce a normalized magnetization $\textstyle{{\bf m}={\bf M}/M_{\rm s}}$, where $\textstyle{M_{\rm s}}$ is the saturation magnetization and $\textstyle{|{\bf m}|=1}$. Moreover, it is convenient to use units where the magnetic field, $M_{\rm s}$, and the magnetic induction are given in Tesla, to connect to the experiment. Then, the Landau-Lifshitz-Gilbert equation has the form:
\begin{equation}\label{LLE}
\displaystyle{\frac{\partial {\bf m}}{\partial t}=-\gamma{\bf m}\times{\bf H}_{\rm eff}+\alpha {\bf m}\times\frac{\partial {\bf m}}{\partial t},}
\end{equation}
where $\textstyle{\gamma}$, $\textstyle{{\bf H}_{\rm eff}}$, and $\textstyle{\alpha}$ are the gyromagnetic ratio, effective magnetic field, and Gilbert damping parameter, respectively. For isotropic ferromagnetic materials, the effective magnetic field is given by (see e.g.~\cite{Landau_9})
\begin{equation} \label{H_eff}
\displaystyle{{\bf H}_{\rm eff}={\bf H}_{\rm ext}+{\bf H}_{\rm d}+D\nabla^2 {\bf m},}
\end{equation}
where $\textstyle{{\bf H}_{\rm ext}}$ and $\textstyle{{\bf H}_{\rm d}}$ are the external and demagnetizing magnetic fields, respectively. The last term in \eqref{H_eff} describes the exchange interaction with exchange stiffness constant $\textstyle{D}$, and $\textstyle{\nabla=(\frac{\partial}{\partial x},\frac{\partial}{\partial y},\frac{\partial}{\partial z})}$.

We stress that both $\textstyle{{\bf m}}$ and $\textstyle{{\bf H}_{\rm d}}$ are spatially inhomogeneous and time-dependent quantities. The connection between them is given by the magnetostatic Maxwell equations~\cite{Landau_9}:
\begin{equation}\label{Maxwell}
\displaystyle{\nabla\times {\bf H}_{\rm d}=0, \phantom{aaaaaaa} \nabla\cdot ({{\bf H}_{\rm d}}+M_{\rm s}{\bf m})=0.}
\end{equation}
The described macroscopic quasi-static approach is valid if the magnon wavelength is larger than the atom spacing, and the magnon phase velocities are smaller than the velocity of light~\cite{Landau_9}. Equations \eqref{LLE}-\eqref{Maxwell} are the main set of nonlinear differential equations for the NG structure. In addition, we use the standard electrodynamic boundary conditions and free boundary conditions for the magnetization, i.e., $\textstyle{\frac{\partial {\bf m}}{\partial {\bf n}}=0}$, because this case describes experiments performed on similar structures~\cite{Langer_2019}.

Equations (3) are mathematically equivalent to the well-known Maxwell electrostatic equations. Therefore, the general solutions of Eqs. (3) can be found using the Green function for the Laplacian, $\textstyle{\nabla^2}$, (see e.g. \cite{Guslienko_2005}). As a result, the demagnetizing field can be written as
\begin{equation} \label{DemagFieldInt}
\displaystyle{{\bf H}_{\rm d}(x,z, t)\!=\!-\frac{M_{\rm s}}{2\pi}\nabla\!\!\int_{-\infty}^{\infty}{\!\!\!\!\!\!dx^{\prime}}\!\!\int_{0}^{l}{\!\frac{m_x(x-x^{\prime})\!+\!m_z(z-z^{\prime})}{(x-x^{\prime})^2\!+\!(z-z^{\prime})^2}dz^{\prime}}.}
\end{equation}
As one can see from Eq.~\eqref{DemagFieldInt}, $\textstyle{m_y}$ does not contribute to the demagnetizing field despite that, in general,  $\textstyle{m_y\neq0}$. Moreover, the $\textstyle{y}$-coordinate is absent in the integrand, as well. This leads us to the general conclusion that $\textstyle{H_{{\rm d},y}}$ is zero for any magnetization direction in the NG. Moreover, the demagnetizing field is zero if the magnetization is parallel to the grooves.
\begin{figure}[t!]
\begin{center}
\includegraphics[width=0.48\textwidth,keepaspectratio]{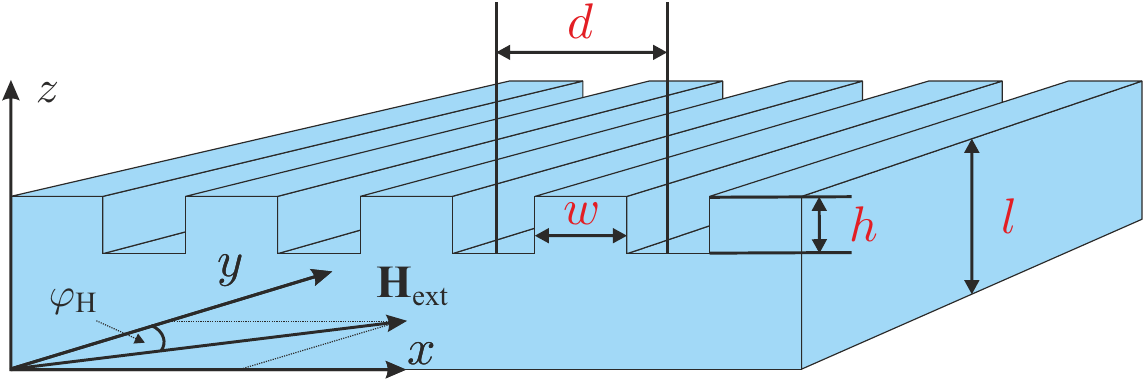}
\end{center}
\vspace{-0.5cm}
\caption{Sketch of the nanograting, the used coordinate system, and the in-plane external magnetic field, $\textstyle{{\bf H}_{\rm ext}}$. }
\end{figure}

The search for the solutions of Eqs.~\eqref{LLE}-\eqref{Maxwell} can be divided into two stages: the steady-state distribution of the magnetization and demagnetizing field and their linear dynamics. In the steady-state, the magnetization, $\textstyle{{\bf m}^0}$, is parallel to $\textstyle{{\bf H}^0_{\rm eff}}$ to minimize the free energy of the ferromagnet~\cite{Landau_9}. In the NG, the spatial distribution of $\textstyle{{\bf H}^0_{\rm eff}}$ is inhomogeneous, as is the spatial distribution of the magnetization (see, e.g., the Eqs.~\eqref{Maxwell}).

For the calculations, we chose polycrystalline Permalloy ($\textstyle{\rm Ni_{80}Fe_{20}}$) as the  NG material. This isotropic ferromagnet possesses very weak magnetostriction, which allows us to exclude from considering the lattice dynamics and consider only magnons. The magnetic parameters~\cite{Langer_2019} for the calculations are the saturation magnetization, $\textstyle{M_{\rm s}=0.9236}$~T and the exchange stiffness, $\textstyle{D=23.6}$~Tnm$^2$. The Gilbert damping parameter $\textstyle{\alpha}$ is fixed to zero for the eigenfrequency analysis. A discussion about different materials and parameters can be found in Sec.~\ref{ValAndApp}. For the geometrical parameters, we will focus on $\textstyle{d\sim 100}$~nm and $\textstyle{l\sim10}$~nm which gives us magnon modes with frequencies $\sim10$~GHz and a first-order magnon mode, which is nonuniform in the $\textstyle{z}$-direction (or $\textstyle{z}$-quantized) and shifted to relatively high frequencies.

\section{Magnetization's steady-state distribution}\label{statics}

The steady-state distribution of the magnetization corresponds to the minimum of the free energy density. In the case of $\textstyle{H_{\rm ext}=0}$, the magnetization is parallel to the grooves of the NG. In the case of $\textstyle{H_{\rm ext}\neq 0}$, the magnetization in the wires tends to be parallel to the grooves, while the magnetization in the film regions (under the patterned volume) tends to be parallel to $\textstyle{{\bf H}_{\rm ext}}$. Thus, the \textit{total} spatial distribution of the steady-state magnetization is non-homogeneous for all directions of $\textstyle{{\bf H}_{\rm ext}}$ except for $\textstyle{\varphi_{\rm H}=0^{\circ}}$.

\begin{figure}[t!]
\begin{center}
\includegraphics[width=0.48\textwidth,keepaspectratio]{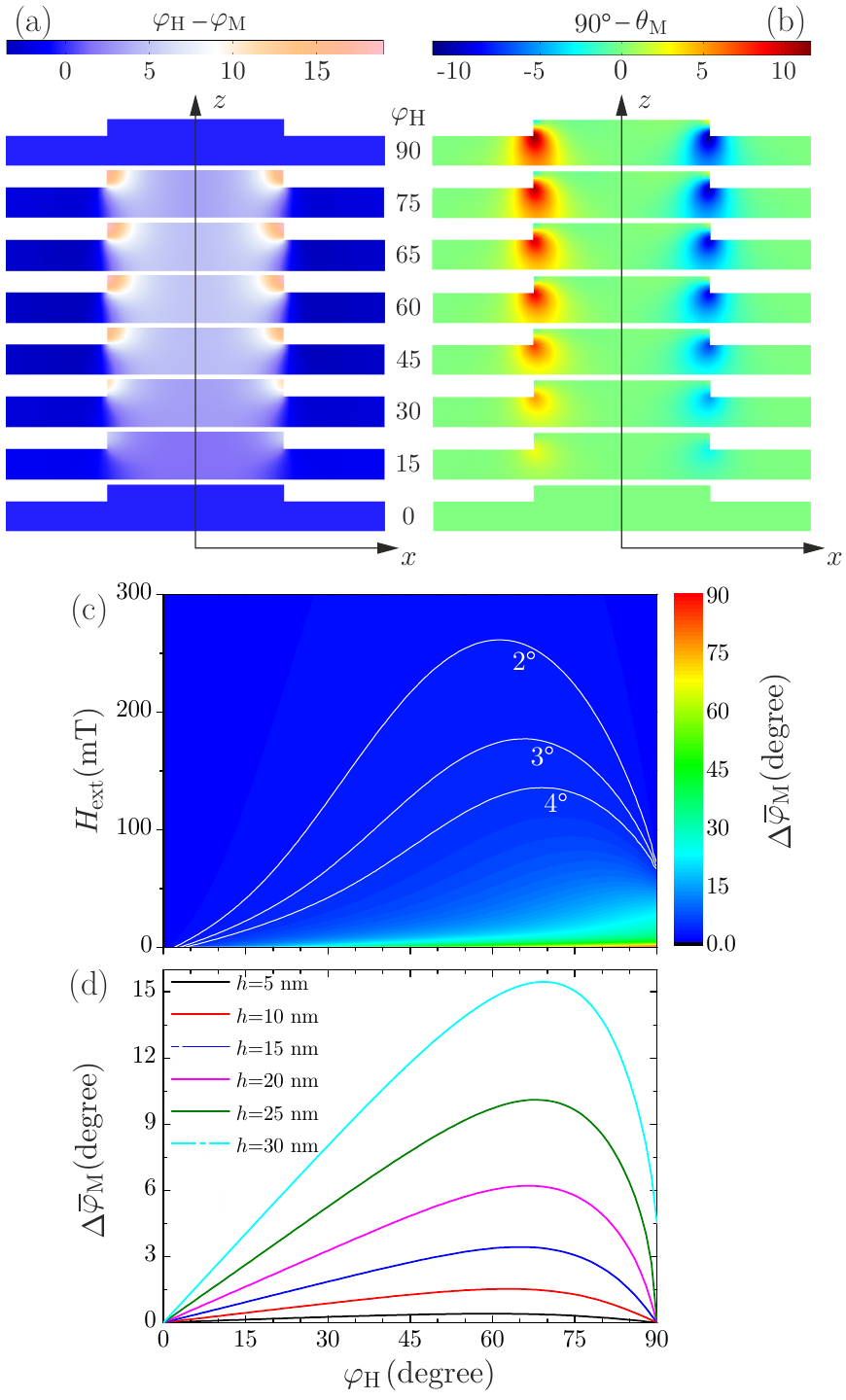}
\end{center}
\vspace{-0.5cm}
\caption{Spatial distributions of (a) $\textstyle{\varphi_{\rm H}-\varphi_{\rm M}}$ and (b) $\textstyle{90^{\circ}-\theta_{\rm M}}$ for different directions of $\textstyle{{\bf H}_{\rm ext}}$ at $\textstyle{H_{\rm ext}=200}$~mT. The values of $\textstyle{\varphi_{\rm H}}$ are shown in between (a) and (b). All numbers are given in degrees. (c) Dependencies of the average deviation angle, $\textstyle{\Delta\overline{\varphi}_{\rm M}}$, on $\textstyle{\varphi_{\rm H}}$ which shows how much the average azimuthal orientation of the magnetization deviates from the $\textstyle{{\bf H}_{\rm ext}}$ direction,  as a function of the direction and strength of $\textstyle{{\bf H}_{\rm ext}}$. The white lines represent the contours of constant angles $\textstyle{\Delta\overline{\varphi}_{\rm M}=2^{\circ}}$, $\textstyle{3^{\circ}}$, and $\textstyle{4^{\circ}}$. (d) Dependencies of $\textstyle{\Delta\overline{\varphi}_{\rm M}}$ on the direction of $\textstyle{{\bf H}_{\rm ext}}$ for different depths of the grooves at $\textstyle{H_{\rm ext}=200}$~mT.}\label{Dev}
\end{figure}

Let us introduce the quantities $\textstyle{\varphi_{\rm M}(x,z)=\arctan{(m_x^0/m_y^0)}}$ and $\textstyle{\theta_{\rm M}(x,z)=\arccos{m_z^0}}$ which are the local azimuthal and polar angles of the magnetization. The deviations of the magnetization direction from the direction of $\textstyle{{\bf H}_{\rm ext}}$ are given by $\textstyle{\varphi_{\rm H}-\varphi_{\rm M}}$ and $\textstyle{90^{\circ}-\theta_{\rm M}}$. The deviations are shown in Fig.~\ref{Dev} for $\textstyle{H_{\rm ext}=200}$ mT and the following fixed geometrical parameters: $\textstyle{d=300}$~nm, $\textstyle{w=140}$~nm, $\textstyle{h=13.2}$~nm, and $\textstyle{l=36.8}$~nm, which are similar to those in recent experiments~\cite{Langer_2019}. At $\textstyle{\varphi_{\rm H}=0}$, the magnetization is uniform, hence $\textstyle{\varphi_{\rm H}-\varphi_{\rm M}=0^{\circ}}$ and $\textstyle{90^{\circ}-\theta_{\rm M}=0^{\circ}}$. At $\textstyle{\varphi_{\rm H}\neq0}$ the azimuthal deviation $\textstyle{\varphi_{\rm H}-\varphi_{\rm M}}$ reaches its maximum of about $\textstyle{19.8^{\circ}}$ at $\textstyle{\varphi_{\rm H}=65^{\circ}}$. The largest azimuthal deviations are localized at the outer corners of the NG wire. At $\textstyle{\varphi_{\rm H}=90^{\circ}}$, the $\textstyle{y}$-component of the magnetization is zero. Thus, the azimuthal deviation is zero. Interestingly, these azimuthal deviations can have both signs. The positive sign means that the magnetization tends to be parallel to the grooves. A small negative azimuthal deviation (up to $\textstyle{-3.5^{\circ}}$) arises in the region under the grooves, where $\textstyle{H^0_{{\rm d},x}}$ is positive (see~\cite{SM} for details). The latter, referred to as a magnetizing field (see, e.g.,~\cite{Langer_2019}), slightly rotates the magnetization towards the $\textstyle{x}$-direction, i.e., perpendicular to the grooves.

The polar deviation of the magnetization, $\textstyle{90^{\circ}-\theta_{\rm M}}$, is shown in Fig. \ref{Dev}(b). In this case, the distribution of the polar deviation is an odd function. This is because the symmetry of the polar and azimuthal deviations (or magnetization components) with respect to the center of the NG wire region corresponds to the symmetry of the $\textstyle{{\bf H}_{\rm d}^0}$ components~\cite{SM}. Note that the positive (negative) polar deviations correspond to positive (negative) $\textstyle{m_z^0}$. It can be seen that such polar deviations increase with increasing $\textstyle{\varphi_{\rm H}}$. The maximum deviation is reached at $\textstyle{\varphi_{\rm H}=90^{\circ}}$. Interestingly, the polar deviations are localized in the inner corners of the NG, while the azimuthal deviations are localized in the outer corners (see Fig 2(a)). This is because the absolute value of the $\textstyle{z}$-component of the demagnetizing field has its maxima at the sharp parts of the NG corners (see~\cite{SM} for details). For the outer corners, $\textstyle{H^0_{{\rm d},z}}$ is localized outside the nanograting structure; for the inner corners, it is localized inside the structure and, thereby, rotates the magnetization in the $z$-direction.

As one can see from the analysis above, the magnetization is not fully saturated (neither homogeneous nor parallel to $\textstyle{{\bf H}_{\rm ext}}$) even at $\textstyle{H_{\rm ext}=200}$~mT, which is strong enough for saturation of a plain Permalloy film. Due to the demagnetizing field, the magnetization in the NG saturates asymptotically with increase of the external magnetic field strength. For example, at $\textstyle{H_{\rm ext}=10}$~T and $\textstyle{\varphi_{\rm H}=90^{\circ}}$, the polar deviation, $\textstyle{90^{\circ}-\theta_{\rm M}}$, changes from $\textstyle{-1.5^{\circ}}$ to $\textstyle{1.5^{\circ}}$. However, the magnetization deviates considerably from the external magnetic field direction only at the corners of the NG. Therefore, for practical purposes, it is convenient to introduce the \textit{average azimuthal angle} of the magnetization, $\textstyle{\overline{\varphi}_{\rm M}=\int{\varphi_{\rm M}dV}}$, which describes the average orientation of the magnetization. Here $\textstyle{dV}$ is the unit volume element. In addition, let us introduce the average deviation angle, $\textstyle{\Delta\overline{\varphi}_{\rm M}=\varphi_{\rm H}-\overline{\varphi}_{\rm M}}$, which describes how much the average azimuthal orientation of the magnetization deviates from the $\textstyle{{\bf H}_{\rm ext}}$ direction. The magnetization can be considered saturated on average if $\textstyle{\sin{\Delta\overline{\varphi}_{\rm M}}\ll1}$.

Figure~\ref{Dev}(c) shows the dependencies of $\textstyle{\Delta\overline{\varphi}_{\rm M}}$ on different magnetic field directions and strengths. The meaning of the white contours in Fig.~\ref{Dev}(c)  is that for any magnetic field direction and strength above a contour $\textstyle{\Delta\overline{\varphi}_{\rm M}=const}$, the values of $\textstyle{\Delta\overline{\varphi}_{\rm M}<const}$. For instance, at $\textstyle{H_{\rm ext}=200}$~mT, $\textstyle{\Delta\overline{\varphi}_{\rm M}}$ is less than $\textstyle{3^\circ}$ at any magnetic field direction.
\begin{figure*}[t!]
\begin{center}
\vspace{-0.4cm}
\includegraphics[width=0.99\textwidth,keepaspectratio]{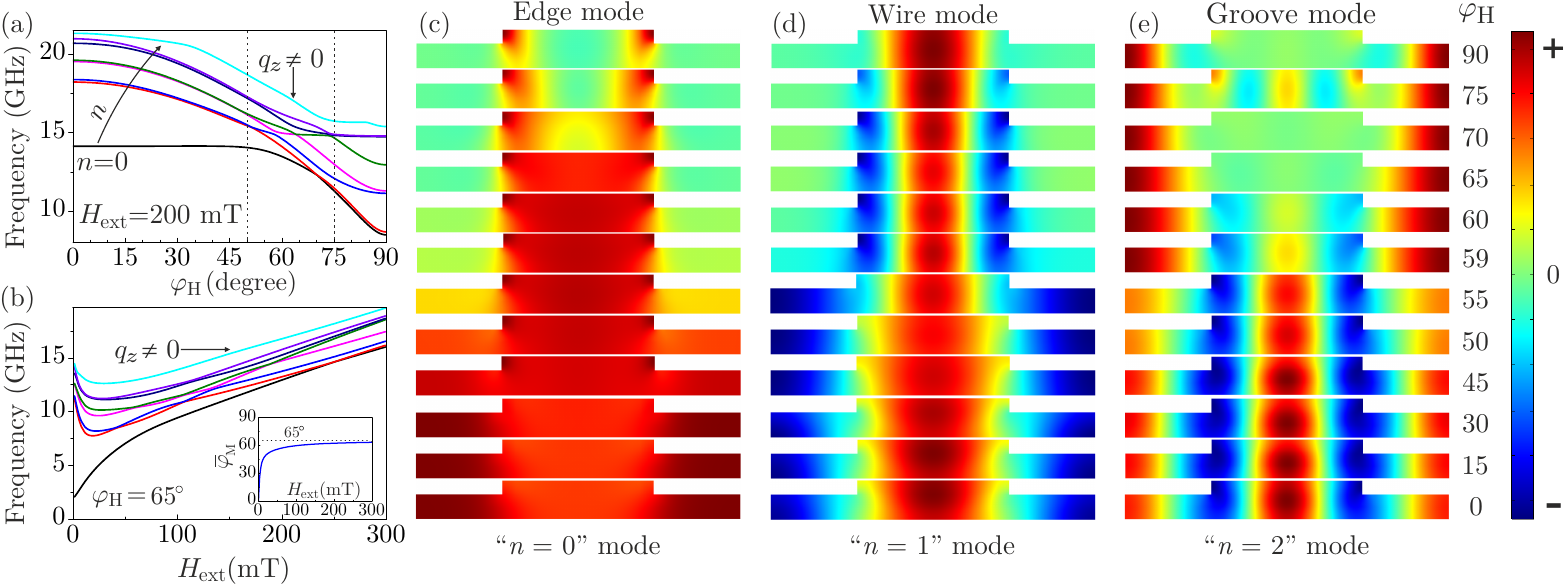}
\end{center}
\vspace{-0.1cm}
\caption{Magnetic field direction (a) and strength (b) dependencies of magnon frequencies. The arrow shows the direction of the number $\textstyle{n}$ increase at $\textstyle{\varphi_{\rm H}\lesssim50^{\circ}}$: the ground magnon branch with $\textstyle{n=0}$ and the three next pairs with, $\textstyle{n=1,2}$, and $3$, respectively. The vertical dashed lines indicate the transition range. The inset in (b) shows the magnetic field dependence of the average azimuthal angle of the magnetization, $\textstyle{\overline{\varphi}_{\rm M}}$. (c)-(e) Evolutions of the main symmetric magnon modes under rotation of the magnetic field. The values of $\textstyle{\varphi_{\rm H}}$ are given in degrees. All quantities here correspond to the center of the Brillouin zone ($\textstyle{k_x=0}$) and $\textstyle{H_{\rm ext}=200}$~mT.}\label{MagDependModEvol}
\end{figure*}

Figure~\ref{Dev}(d) shows the dependencies of $\textstyle{\Delta\overline{\varphi}_{\rm M}}$ on the external magnetic field direction for different depths of the grooves, $\textstyle{h}$, calculated at $\textstyle{H_{\rm ext}=200}$~mT. The maximum position slightly shifts toward higher $\textstyle{\varphi_{\rm H}}$ from $\textstyle{60^{\circ}}$ for the shallow NG ($\textstyle{h=5}$~nm) to $\textstyle{69^{\circ}}$ for the deep NG ($\textstyle{h=30}$~nm). The reason why $\textstyle{\Delta\overline{\varphi}_{\rm M}}$ demonstrates a slow increase at $\textstyle{\varphi_{\rm H}\lesssim50^{\circ}}$ and a rapid decrease at $\textstyle{\varphi_{\rm H}\gtrsim75^{\circ}}$ is due to the collective dipole-dipole interaction. That is, the spatial region, where the magnetization direction is close to the field direction grows with increasing $\textstyle{H_{\rm ext}}$.  At $\textstyle{\varphi_{\rm H}}$, which gives the maximum of $\textstyle{\Delta\overline{\varphi}_{\rm M}}$, this region becomes large enough to change the direction of the magnetization in the wire region of the NG. Then, the magnetization in this region attracts the magnetization at the corners of the NG leading to the rapid decrease of $\textstyle{m_y}$. As a result, $\textstyle{\Delta\overline{\varphi}_{\rm M}}$ drops sharply to zero at $\textstyle{\varphi_{\rm H}=90^{\circ}}$. Note, that for the NG with $\textstyle{h=30}$~nm, one must apply much larger magnetic fields than $\textstyle{H_{\rm ext}=200}$~mT to saturate the magnetization for any $\textstyle{\varphi_{\rm H}}$.

Keeping this physical picture in mind, we can turn to the analysis of magnons in the NG.

\section{Classification of the magnon eigenmodes}\label{dynamics}

The magnetic dynamics of a ferromagnet consists of the spatially inhomogeneous precession of the magnetization around $\textstyle{{\bf H}^0_{\rm eff}}$. This time- and space-dependent precessional motion can be considered as a superposition of the magnon eigenmodes of the NG. To describe it, we have linearized the main set of equations by introducing a dynamic magnetization, $\textstyle{\delta{\bf m}}$, with $\textstyle{\delta m\ll m^0}$, and a dynamic demagnetizing field, $\textstyle{\delta{\bf H}_{\rm d}}$, with $\textstyle{\delta H_{\rm d}\ll M_{\rm s}}$. In order to calculate the dispersion curves and spatial profiles of the magnon modes, we performed an eigenfrequency analysis with Floquet-Bloch periodic boundary conditions, i.~e. $\textstyle{{\bf u}_{\rm d}={\bf u}_{\rm s}}\exp{(-i k_x d )}$, where $\textstyle{k_x}$ is the Bloch wavenumber, and $\textstyle{{\bf u}_{\rm d}}$ as well as $\textstyle{{\bf u}_{\rm s}}$ are dependent variables at the destination and source boundaries, respectively.

The solutions of the eigenfrequency problem give the spatially inhomogeneous complex-valued Fourier components of the dynamic magnetization and demagnetizing field. For the characterization of the magnon modes' spatial profiles, let us choose the $\textstyle{z}$-component of the dynamic magnetization because it is nonzero for any in-plane $\textstyle{{\bf H}_{\rm ext}}$. Below, we focus on the spatial profiles, which correspond to \textit{the center} of the Brillouin zone, i.e., $k_x=0$. The real-valued solution for $\textstyle{\delta m_z}$ in the time domain can be expressed in the form $\textstyle{\delta m_z=|\delta m_{z,\omega}|cos(\omega t+\phi_z)}$, where $\textstyle{\omega}$ is the magnon angular frequency, $\textstyle{\delta m_{z,\omega}}$ is the Fourier component of $\textstyle{\delta m_z}$, and $\textstyle{\phi_z=\text{atan}2(\text{Im}(\delta m_{z,\omega}),\text{Re}(\delta m_{z,\omega}))}$ is the phase. The latter is spatially inhomogeneous. Thus, the magnon modes' spatial profiles change during propagation~\cite{Langer_2019}. Moreover, the difference between the supremum and infimum values of $\textstyle{\delta m_z}$, $\textstyle{{\cal D}_{\rm m}(t)=\sup(\delta m_z)-\inf(\delta m_z)}$, varies with time. However, we found that at the time, which corresponds to the maximum of $\textstyle{{\cal D}_{\rm m}(t)}$, each mode profile becomes purely symmetric or antisymmetric with respect to the center of the NG wire region. This time can be found as $\textstyle{t=(\tau_s-\phi_{z,c})/\omega}$, where $\textstyle{\tau_s=s \pi}$ for symmetric modes ($\textstyle{s}$ is an integer number), $\textstyle{\tau_s=s \pi/2}$ for antisymmetric modes, and $\textstyle{\phi_{z,c}}$ is the value of the phase at the center of the unit cell of the NG ($\textstyle{x=0, z=l/2}$ in our coordinate system). Thus, we use the symmetry at this time for characterizing a magnon mode. This symmetry determines the possibility of the mode being excited by a spatially homogeneous excitation (the laser pulse or the uniform ac- magnetic field in a FMR experiment).

According to the Floquet-Bloch theorem, each eigenmode contains an infinite number of wave vectors $\textstyle{q_x=k_x+n\,\frac{2\pi}{d}}$, where $\textstyle{n}$ is an integer number. To characterize the magnon modes by the number, $\textstyle{n}$, we perform the spatial Fourier transform
\begin{equation} \label{FourierTransform}
\displaystyle{\delta m_{z,n}=\int{\delta m_z(x,z) \exp{\left(in \frac{2\pi}{d}x\right)}\exp{(i q_z z)}dV}}
\end{equation}
and find the set of $\textstyle{n}$, which dominate in each eigenmode. Here $\textstyle{q_z}$ is the z-component of the wavevector. Further we will focus on the quasiuniform eigenmodes in the $z$ direction and put $\textstyle{q_z=0}$.

The result of an eigenfrequency analysis is presented in Fig.~\ref{MagDependModEvol}(a), which shows the angle dependence of the magnon frequencies, $\omega/2\pi$ at $\textstyle{H_{\rm ext}=200}$~mT. Here, one can observe a complicated non-monotonic behavior with magnon-magnon interaction and corresponding avoided crossings of the interacting modes at the intersections. Despite that, the general tendency is typical for ferromagnets (see, for instance,~\cite{Scherbakov_2019, Khokhlov_2019}): the magnon frequency decreases with changing of the direction of $\textstyle{{\bf H}_{\rm ext}}$ from the easy axis($\textstyle{\varphi_{\rm H}=0^{\circ}}$) to the hard axis ($\textstyle{\varphi_{\rm H}=90^{\circ}}$). Interestingly, the frequency of the ground magnon mode does not significantly depend on the direction of $\textstyle{{\bf H}_{\rm ext}}$ for $\textstyle{\varphi_{\rm H}\lesssim50^{\circ}}$ and can be estimated with reasonable accuracy using the Kittel formula~\cite{Lisenkov_2015}.

Figure~\ref{MagDependModEvol}(b) shows the magnetic field dependence of the magnon frequencies at $\textstyle{\varphi_{\rm H}=65^{\circ}}$. The ground magnon branch (black line) demonstrates a simple Kittel-like behavior. All following branches firstly decrease and, then, increase in their frequencies with increasing external magnetic field strength. The turning point at $H_{\rm ext}\approx 15$ mT corresponds to the field strength overcoming the anisotropy field when the steady-state magnetization turns toward $\textstyle{{\bf H}_{\rm ext}}$. This is illustrated in the inset, which shows the magnetic field dependence of the average azimuthal angle of the magnetization, $\textstyle{\overline{\varphi}_{\rm M}(H_{\rm ext})}$, for the chosen $\varphi_{\rm H}$.

The spatial profiles of the magnon modes in the DE and BV geometries can be found in~\cite{Langer_2019} and~\cite{SM}. For the DE geometry, the magnon modes (quasiuniform in the $z$ direction) can be completely characterized by the single spatial harmonic along the x-axis, characterized by the number $\textstyle{n}$. The ground quasiuniform magnon mode corresponds to $\textstyle{n=0}$, and the next pair of antisymmetric and symmetric modes have $\textstyle{n=1}$ (see the amplitudes of the spatial Fourier transform, $\textstyle{|\delta m_{z,n}|}$, in ~\cite{SM}), etc. In the BV geometry, the localization of magnon modes is fragmented into separate regions of wires and grooves. As a result, each magnon mode consists of several dominant $\textstyle{n}$~\cite{TwoNumbers}. For example, the ground symmetric edge mode possesses three dominant $\textstyle{n}$: $\textstyle{n=3, 2, 0}$ (hereafter, the sequence is given in descending order of $\textstyle{|\delta m_{z,n}|}$, see~\cite{SM}). The symmetric wire mode can be characterized by $\textstyle{n=1, 0, 2}$, and the symmetric groove mode by $\textstyle{n=2, 0, 1}$.

Figures \ref{MagDependModEvol}(c)-(e) show the evolution of the spatial profiles of the three lowest symmetric magnon modes under rotation of the magnetic field. One can see that the ground quasiuniform (Kittel) mode ($\textstyle{n=0}$) in the DE geometry evolves to the edge mode in the BV geometry. The magnon modes with $\textstyle{n=1}$ and $\textstyle{n=2}$ evolve to wire and groove modes, respectively. Note that we add the field direction $\textstyle{\varphi_{\rm H}= 59^{\circ}}$ to show the smooth transition of the $\textstyle{n=2}$ mode to the groove mode. One can see that for the field directions up to $\textstyle{\approx 50^{\circ}}$, the magnon modes can be considered DE-like, because their spatial profiles maintain their distributions and can be characterized by the single number $\textstyle{n}$. Similarly, starting from $\textstyle{\varphi_{\rm H}\approx 75^{\circ}}$ and up to $\textstyle{90^{\circ}}$, the magnon modes can be considered BV-like. For the range $\textstyle{50^{\circ}-75^{\circ}}$, the magnon modes' profiles considerably differ from DE-like and BV-like modes. We call these modes \textit{transition magnon modes}. Note that the spatial modulation in the wire region of the groove mode at $\textstyle{\varphi_{\rm H}= 75^{\circ}}$ is due to coupling with the higher order wire magnon mode (similar to the 5-th mode in Fig. S6(d) in~\cite{SM}).

Figures \ref{MagDependModEvol}(c)-(e) clearly show that the intermediate directions of $\textstyle{{\bf H}_{\rm ext}}$ can be used for adjusting the spatial overlap of the selected magnon mode with another periodic excitation, such as an elastic wave, to maximize the magneto-elastic coupling strength~\cite{Bombeck_2012,Verba_2018,Yang_2019,Godejohann_2020,Babu_2021}. In the latter case, the magneto-elastic interaction can occur between symmetric magnon and symmetric phonon modes, as well as between symmetric magnon modes and antisymmetric phonon modes~\cite{Godejohann_2020} or in other combinations~\cite{Bombeck_2012,Verba_2018}.

\begin{figure}[t!]
\begin{center}
\includegraphics[width=0.48\textwidth,keepaspectratio]{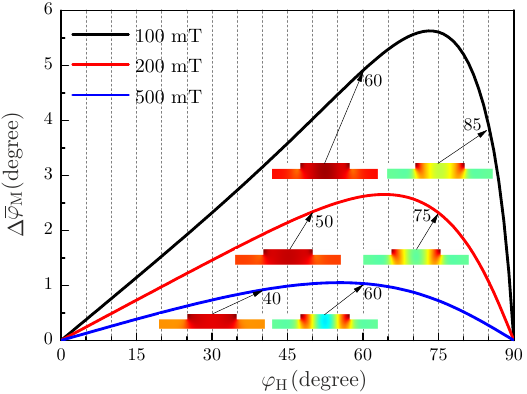}
\end{center}
\vspace{-0.5cm}
\caption{Magnetic field direction dependence of $\textstyle{\Delta\overline{\varphi}_{\rm M}}$, showing the transition from DE-like modes to BV-like modes for three different values of $\textstyle{H_{\rm ext}}$.}\label{Transitions}
\end{figure}

Notably, the transition range is defined by the steady-state magnetization distribution in the NG. From the analysis of the previous section and, in particular, from Fig.~\ref{Dev}, one can see that the transition range is located around the maximum of $\textstyle{\Delta\overline{\varphi}_{\rm M}}$. We stress that the low-order magnon modes are influenced by the static demagnetizing fields more strongly than the high-order modes because the wavelengths of the high-order modes are smaller than the NG period. Figure~\ref{Transitions} shows how the transition range depends on the external magnetic field strength: the maximum of $\textstyle{\Delta\overline{\varphi}_{\rm M}}$ shifts towards its asymptotic limit of $\textstyle{45^{\circ}}$ with increasing $\textstyle{H_{\rm ext}}$. In particular, for $\textstyle{\varphi_{\rm H}=60^{\circ}}$, the ground symmetric magnon mode is still the quasiuniform mode at $\textstyle{H_{\rm ext}=100}$~mT, but becomes already the edge mode at $\textstyle{H_{\rm ext}=500}$~mT. The position of the transition range depends on the interplay of the external magnetic field and the demagnetizing field. The transition range width decreases with the increasing field strength. For commonly used magnetic fields ($\textstyle{\sim 100}$~mT), the modes are DE-like at $\textstyle{\varphi_{\rm H}<= 45^{\circ}}$. The existence of the transition magnon modes in the transition range of the magnetic field directions with their relation to the static NG magnetization is one of the main results of the paper.

\begin{figure}[t!]
\begin{center}
\includegraphics[width=0.45\textwidth,keepaspectratio]{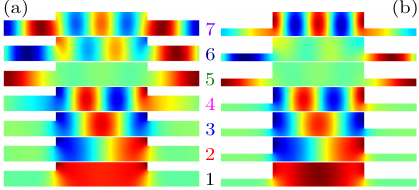}
\end{center}
\caption{Spatial distributions of transition magnon modes for two depths of the grooves: $\textstyle{h=13.2}$~nm and $\textstyle{h=25}$~nm. The colors of sequence numbers correspond to the line colors in Fig.~~\ref{MagDependModEvol}(a), (b), and the color scale is the same as in Fig.~\ref{MagDependModEvol}~(c)-(e).}\label{TransitionModes}
\end{figure}

Spatial profiles of transition magnon modes are shown in Fig.~\ref{TransitionModes} for $\textstyle{\varphi_{\rm H}= 65^{\circ}}$ and two different depths of the grooves. A distinctive feature of these modes is that they are well distributed in separate parts of the NG. For example, the ground transition mode, which transforms from the uniform to the edge mode, is localized in the wire region. However, in contrast to the edge mode, it has there a quasiuniform distribution (see Fig~\ref{MagDependModEvol}(c)). This mode is also different from the wire mode because it has a significantly different spatial spectrum given by $\textstyle{n=0,1,3}$, while for the wire mode $\textstyle{n=1,0,2}$ (see Fig. S4 in~\cite{SM}). All four lowest magnon modes are localized in the wire region, and their sequence number indicates the number of peaks in the wire region. Then come the symmetric and antisymmetric transition modes localized in the grooves, but with a broader spatial distribution than the groove mode in the BV geometry. The following modes of higher orders, as well as the high-order modes in the DE and BV geometries, are weakly influenced by the demagnetizing field because their wavelengths are smaller than the NG period. The spatial profiles of transition magnon modes weakly depend on the depth of the grooves.

\section{Magnon dispersion and nonreciprocity}\label{dispersion}
\begin{figure*}[t!]
\begin{center}
\vspace{-0.4cm}
\includegraphics[width=0.95\textwidth,keepaspectratio]{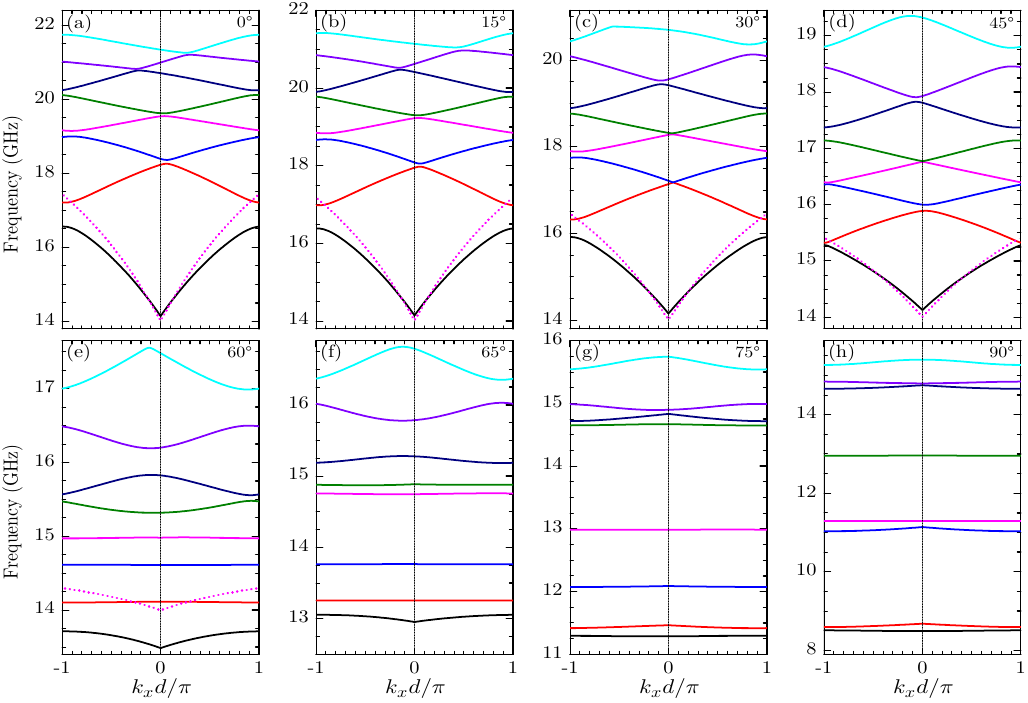}
\end{center}
\vspace{-0.1cm}
\caption{(a)-(h) Magnon dispersion evolution under rotation of $\textstyle{{\bf H}_{\rm ext}}$ from $\textstyle{\varphi_{\rm H}=0^{\circ}}$ (DE) to $\textstyle{\varphi_{\rm H}=90^{\circ}}$ (BV) at $\textstyle{H_{\rm ext}=200}$~mT. The values of $\textstyle{\varphi_{\rm H}}$ are shown in the right upper corners. Note  that the vertical scales are different in (a)-(h). The pink dotted lines represent the corresponding thin film dispersions (see Eq. \eqref{FilmDisp}).}\label{Dispersions}
\end{figure*}

The dispersion is another crucial characteristic of magnons that gives the wave vector dependencies of the magnon modes' frequencies and their group velocities, as well as the spectral positions and widths of the magnon band gaps. It can be directly measured by Brillouin light scattering techniques~\cite{Gubbiotti_2005,Gubbiotti_2007,Kostylev_2008,Krawczyk_2014,Gubbiotti_2021} and determines the transient signals in the magneto-optical pump-probe experiments with microscopic resolution.

Figures~\ref{Dispersions}~(a)-(h) show the evolution of the magnon dispersions under rotation of the magnetic field from $\textstyle{\varphi_{\rm H}=0^{\circ}}$ to $\textstyle{\varphi_{\rm H}=90^{\circ}}$ at $\textstyle{H_{\rm ext}=200}$~mT. The nonreciprocity and indirect band gaps can be observed for all directions of $\textstyle{{\bf H}_{\rm ext}}$ except $\textstyle{\varphi_{\rm H}=90^{\circ}}$. With increasing $\textstyle{\varphi_{\rm H}}$, the magnon dispersion branches shift to lower frequencies and several low-lying branches become flat. Due to the vanishingly small dynamic demagnetizing field of the ground DE-like branch, it can be qualitatively described by the thin-film dispersion~\cite{Kalinikos_1986, Gallardo_2014}:
\begin{eqnarray} \label{FilmDisp}
\displaystyle{\omega^2=\gamma\left(H_{\rm ext}+M_{\rm s}\left(1-\frac{1-e^{-q_xl}}{q_xl}\right)\cos^2{\!\varphi_{\rm H}}+Dq_x^2\right)},
\nonumber\\
\displaystyle{\times\gamma\left(H_{\rm ext}+M_{\rm s}\left(\frac{1-e^{-q_xl}}{q_xl}\right)+Dq_x^2\right),\phantom{abcdefghi}}
\end{eqnarray}
in the range of magnetic field directions before the transition range, i.e., at $\textstyle{\varphi_{\rm H}\lesssim50^{\circ}}$. This dispersion is shown by dots in Fig.~\ref{Dispersions}(a)-(e) in the first Brillouin zone. At $\textstyle{\varphi_{\rm H}}=0^{\circ}$, the seven lowest branches originate from the Eq.~\eqref{FilmDisp} due to the folding of the dispersion curves. Then comes the $\textstyle{z}$-quantized branch \#8, which starts its own series of high frequency folded dispersion curves. The agreement between the numerical solutions and Eq.\eqref{FilmDisp} becomes unfulfilled in the transition range and the range of BV-like modes for the chosen set of parameters because of relatively strong demagnetizing fields. Therefore, a better agreement can be reached if $\textstyle{q_x l\ll1}$ and $\textstyle{h/l\ll1}$. Note, that in the geometry close to BV the dispersions of several lowest branches are flat~\cite{Gallardo_2018} already at $\textstyle{h=5}$~nm (all other parameters are fixed as in Sec.~\ref{Eqs}).

A fundamental characteristic of the dispersion curves is the nonreciprocity, i.e., $\textstyle{\omega(k_x)\neq\omega(-k_x)}$. The nonreciprocity in the NG arises from the symmetry of the structure and the dynamical demagnetizing fields. In the DE geometry, both $\textstyle{\delta m_x}$ and $\textstyle{\delta m_z}$ contribute equally to $\textstyle{\delta{\bf H}_{\rm d}}$ (see Eq.~\eqref{DemagFieldInt}). As a result, at a random time, $\textstyle{\delta{\bf H}_{\rm d}}$ is neither symmetric nor antisymmetric, which causes the propagating behavior of the modes. In the BV geometry, $\textstyle{\delta{\bf H}_{\rm d}}$ is defined by $\textstyle{\delta m_z}$. As a result, the parity of the modes is conserved at any time and the dispersion is reciprocal.

\begin{figure}[t!]
\begin{center}
\includegraphics[width=0.4\textwidth,keepaspectratio]{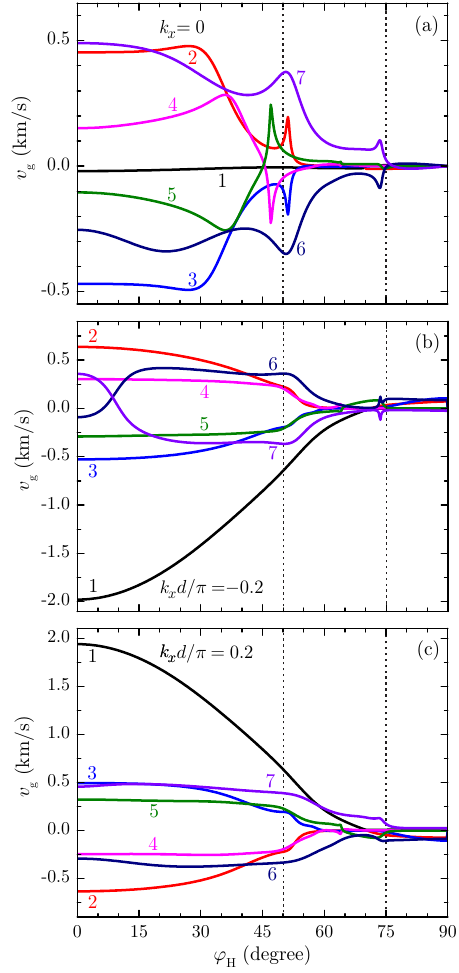}
\end{center}
\vspace{-0.25cm}
\caption{Group velocities of the seven lowest magnon modes for $\textstyle{k_x=0}$ (a), $\textstyle{k_xd/\pi=-0.2}$ (b) and $\textstyle{k_xd/\pi=0.2}$ (c). The vertical dashed lines indicate the transition range.}\label{Vg}
\end{figure}

One may note, that the slopes of the dispersion curves depend on $\textstyle{\varphi_{\rm H}}$ non-monotonically. To show this explicitly, we calculated the group velocities $\textstyle{v_{\rm g}=\frac{d\omega}{d k_x}}$ at the center of the Brillouin zone and at two fixed values of the Bloch wavevector $\textstyle{k_x d/\pi=\pm0.2}$. The dependencies of $\textstyle{v_{\rm g}(\varphi_{\rm H})}$ for the seven lowest magnon modes are shown in Fig.~\ref{Vg}. First, let us consider the case $\textstyle{k_x=0}$. The group velocity of the ground quasiuniform magnon mode \#1 is close to zero (Fig.~\ref{Vg}(a)). This is due to the vanishingly small dynamic demagnetizing fields that the quasiuniform mode creates. Consequently, this mode is reciprocal and has vanishingly small $\textstyle{v_{\rm g}}$ at $\textstyle{k_x=0}$. The next pairs, namely \#2 and \#3, \#4 and \#5, \#6 and \#7, possess similar dependencies with opposite signs. For $\textstyle{\varphi_{\rm H}=90^{\circ}}$, the group velocities for all modes are zero. Note that the highest magnon group velocity is observed not for the DE geometry. The magnon mode \#3 with $\textstyle{n=1}$ reaches $\textstyle{v_{\rm g}\approx-0.5}$~km/s around $\textstyle{\varphi_{\rm H}=30^\circ}$. The group velocities of modes \#2 -- 7 rapidly decrease at the angles $\textstyle{\varphi_{\rm H}\gtrsim 50^{\circ}}$, which corresponds to the transition range. Thus, the group velocities of transition modes are close to zero. The lower magnon modes are more strongly influenced by the static demagnetizing fields than the higher-order modes, as the wavelength of these modes is smaller than the NG period. Therefore, the group velocities of high order modes are nonzero up to $\textstyle{\varphi_{\rm H}=90^{\circ}}$.

At $\textstyle{k_x\neq0}$, the fastest magnon mode is the ground one in the DE geometry with $\textstyle{|v_{\rm g}|\approx 2}$~km/s. The pairs \#2 and \#3, \#4 and \#5, \#6 and \#7, possess similar dependencies with opposite signs as in the case $\textstyle{k_x=0}$. The velocities of transition modes are close to zero due to the flattening of the dispersions. Interestingly, the indirect band gap between the \#6 and \#7  modes at $\textstyle{\varphi_{\rm H}\approx 10^{\circ}}$ causes the strong dependence of the group velocity on the magnon propagation direction: the group velocity of mode \#7 changes from zero to $\textstyle{0.5}$~km/s for negative and positive propagation directions, respectively.

\section{Validity and applications}\label{ValAndApp}

Finally, let us discuss the validity and generality of the obtained numerical results. The validity of our approach was successfully proven by solving standard micromagnetic problems~\cite{SMMP,SM}. Our results are valid for relatively thin NGs, i.e. if the $\textstyle{z}$-quantized magnon branch is well above the ground magnon branch. We found numerically that both the magnon dispersions and spatial profiles remain qualitatively similar for surface modulated magnonic crystals with parameters that satisfy the criterion: $\textstyle{l_{\rm e}d/l^2>c_0}$, where $\textstyle{l_{\rm e}=\sqrt{D/M_{\rm s}}}$ is the exchange length, and $\textstyle{c_0\approx 1}$. For the parameters used in the paper, $\textstyle{l_{\rm e}d/l^2\approx1.1}$ which is close to the lower limit of $\textstyle{l_{\rm e}d/l^2}$. The exchange length is of the same order for most common ferromagnetic metals. For example, for $\textstyle{\rm CoFeB}$: $\textstyle{l_{e\rm }\approx4.7}$~nm (see e.g.~\cite{CoFeB_le}), for permalloy: $\textstyle{l_{e\rm }\approx5.1}$~nm, for nickel: $\textstyle{l_{e\rm }\approx7.7}$~nm (see e.g.~\cite{Ni_le}), and for cobalt: $\textstyle{l_{e\rm }\approx8.1}$~nm (see e.g.~\cite{Co_le}). For such materials as $\textstyle{\rm CoFe}$: $\textstyle{l_{\rm e}\approx3.2}$ (see e.g.~\cite{Piao_2011}) and for FeGa: $\textstyle{l_{e\rm }\approx4.1}$~nm (see e.g. ~\cite{Gopman_2017}), the above criterion is violated for $\textstyle{l>30}$~nm. For the mentioned dielectric ferromagnet YIG, the exchange length is larger than in metals: $\textstyle{l_{e\rm }\approx17.6}$~nm (see~\cite{Klingler_2014}). Note, that in addition to the mentioned criterion, the magnon characteristics can be considerably different for the extremal geometrical parameters. For instance, the edge mode does not exist if $\textstyle{h/l\rightarrow 0}$ and the groove mode, obviously, does not exist in full magnonic crystals ($\textstyle{h\approx l}$).  In the case $\textstyle{l_{\rm e}d/l^2<1}$, the transition region is existent as far as demagnetizing fields exist. It is located also between $\textstyle{\varphi_{\rm H}=45^{\circ}}$ and $\textstyle{90^{\circ}}$. However, with increasing film thickness, the slope of the ground Kittel-like magnon branch increases and the $\textstyle{z}$-quantized branch downshifts. Therefore, eventually, these branches intersect and the corresponding magnon mode characteristics become considerably modified due to strong magnon-magnon coupling which causes a band structure renormalization. The detailed description of this case is beyond the scope of this paper.

The presence of the transition magnon modes with complex characteristics is important for experimental studies and technological applications based on the excitation and detection of coherent magnons by ultrashort laser pulses. Such investigations are typically implemented for intermediate directions of the external magnetic field. The choice of the experimental geometry is determined by the mechanism of excitation: ultrafast thermal- and/or strain- (excited by thermal stress) induced modulations of the magnetic anisotropy, which are the most effective at the intermediate directions of $\textstyle{{\bf H}_{\rm ext}}$. The coherent magnon response in NG is a superposition of the excited magnon modes. The excitation efficiency of a certain mode is determined by its spatial overlap with the optically induced impact~\cite{VanKampen_2002,Bombeck_2012,Scherbakov_2019}. Control of the spatial profiles of the magnon modes by the direction and/or strength of the external magnetic field allows tuning of the spatial distribution and frequency spectrum of the coherent magnon response. This approach can be especially effective in a NG with enhanced magneto-elastic interaction, in which the excitation of coherent magnons is mediated by optically excited phonon eigenmodes of the NG~\cite{Salasyuk_2018,Godejohann_2020}. The interaction of specific phonon and magnon eigenmodes can be fully suppressed in the case of poor spatial matching, or enhanced up to formation of a hybridized state in the case of perfectly matched modes~\cite{Godejohann_2020,Babu_2021}. This tunability is important for hybrid magnonics~\cite{Awschalom_2021,Li-2020} with potential quantum applications.

Optical excitation of the propagating spin waves is another possible experimental route~\cite{Ivanov_2012,Jackl_2017,Khokhlov_2019}, which can utilize the tunability of transition magnon modes. Nonreciprocity of the magnon modes makes possible the excitation of propagating spin waves even in the case of uniform optical excitation. However, the magnon group velocities at intermediate directions are pretty low for the considered geometrical/material parameters. The typical value of the Gilbert damping for Permalloy is $\textstyle{\alpha=0.01}$ which corresponds to a magnon lifetime of $\textstyle{\sim 1}$~ns. In this case, the propagation length of a magnon mode with $\textstyle{v_g=0.5}$~km/s  is $\textstyle{\sim 500}$~nm, which is comparable with the NG period. To increase the propagation distance, one can use materials with weak damping, e.g. yttrium-iron-garnet (YIG) or $\textstyle{\rm Co_{0.25}Fe_{0.75}}$ with  comparable Gilbert damping $\textstyle{\sim10^{-4}}$~\cite{Serga_2010,Schoen_2016}. The group velocities also increase with increasing ferromagnet film thickness~\cite{Yu_2012}. For example, our calculations give $\textstyle{v_{\rm g}>15}$~km/s for the ground magnon mode near the center of the Brillouin zone in case of $\textstyle{l=200}$~nm (all other parameters taken the same as in Sec.~\ref{Eqs}).

\section{CONCLUSION}\label{conclusion}
To conclude, we have investigated the magnon modes of ferromagnetic nanogratings at intermediate in-plane directions of the external magnetic field that were not addressed so far. We have used COMSOL Multiphysics to calculate the magnon dispersions, spatial profiles, and their dependencies on the direction and strength of the external magnetic field. We have found a transition range of magnetic fields with a width of $\textstyle{\sim25^{\circ}}$ where the magnon characteristics are very different from the characteristics in the widely investigated Damon-Eshbach and Backward Volume geometries. The range width and its position depend on the interplay of the external magnetic field and the static demagnetizing field. The modes in the transition range are slowly propagating waves that are well separated in certain parts of the NG because of their unique combinations of spatial Fourier harmonics. We have shown that by changing the magnetic field direction around the transition range, one can switch between three different types of magnon modes and choose the magnon modes of desirable characteristics.

\section{ACKNOWLEDGEMENTS}

We acknowledge the late B. A. Glavin for his valuable contribution to this work. We are thankful to Tetiana Linnik and Oksana Chubykalo-Fesenko for fruitful discussions. The work was supported by the Bundesministerium f\"{u}r Bildung und Forschung through the project VIP+ "Nanomagnetron" and by the Volkswagen Foundation (grant no. 97758).

\bibliography{Bibliography}

\begin{thebibliography}{72}%
\makeatletter
\providecommand \@ifxundefined [1]{%
 \@ifx{#1\undefined}
}%
\providecommand \@ifnum [1]{%
 \ifnum #1\expandafter \@firstoftwo
 \else \expandafter \@secondoftwo
 \fi
}%
\providecommand \@ifx [1]{%
 \ifx #1\expandafter \@firstoftwo
 \else \expandafter \@secondoftwo
 \fi
}%
\providecommand \natexlab [1]{#1}%
\providecommand \enquote  [1]{``#1''}%
\providecommand \bibnamefont  [1]{#1}%
\providecommand \bibfnamefont [1]{#1}%
\providecommand \citenamefont [1]{#1}%
\providecommand \href@noop [0]{\@secondoftwo}%
\providecommand \href [0]{\begingroup \@sanitize@url \@href}%
\providecommand \@href[1]{\@@startlink{#1}\@@href}%
\providecommand \@@href[1]{\endgroup#1\@@endlink}%
\providecommand \@sanitize@url [0]{\catcode `\\12\catcode `\$12\catcode
  `\&12\catcode `\#12\catcode `\^12\catcode `\_12\catcode `\%12\relax}%
\providecommand \@@startlink[1]{}%
\providecommand \@@endlink[0]{}%
\providecommand \url  [0]{\begingroup\@sanitize@url \@url }%
\providecommand \@url [1]{\endgroup\@href {#1}{\urlprefix }}%
\providecommand \urlprefix  [0]{URL }%
\providecommand \Eprint [0]{\href }%
\providecommand \doibase [0]{https://doi.org/}%
\providecommand \selectlanguage [0]{\@gobble}%
\providecommand \bibinfo  [0]{\@secondoftwo}%
\providecommand \bibfield  [0]{\@secondoftwo}%
\providecommand \translation [1]{[#1]}%
\providecommand \BibitemOpen [0]{}%
\providecommand \bibitemStop [0]{}%
\providecommand \bibitemNoStop [0]{.\EOS\space}%
\providecommand \EOS [0]{\spacefactor3000\relax}%
\providecommand \BibitemShut  [1]{\csname bibitem#1\endcsname}%
\let\auto@bib@innerbib\@empty
\bibitem [{\citenamefont {Lenk}\ \emph {et~al.}(2011)\citenamefont {Lenk},
  \citenamefont {Ulrichs}, \citenamefont {Garbs},\ and\ \citenamefont
  {M\"{u}nzenberg}}]{Lenk_2011}%
  \BibitemOpen
  \bibfield  {author} {\bibinfo {author} {\bibfnamefont {B.}~\bibnamefont
  {Lenk}}, \bibinfo {author} {\bibfnamefont {H.}~\bibnamefont {Ulrichs}},
  \bibinfo {author} {\bibfnamefont {F.}~\bibnamefont {Garbs}},\ and\ \bibinfo
  {author} {\bibfnamefont {M.}~\bibnamefont {M\"{u}nzenberg}},\ }\bibfield
  {title} {\bibinfo {title} {The building blocks of magnonics},\ }\href
  {https://doi.org/https://doi.org/10.1016/j.physrep.2011.06.003} {\bibfield
  {journal} {\bibinfo  {journal} {Phys. Rep.}\ }\textbf {\bibinfo {volume}
  {507}},\ \bibinfo {pages} {107 } (\bibinfo {year} {2011})}\BibitemShut
  {NoStop}%
\bibitem [{\citenamefont {Krawczyk}\ and\ \citenamefont
  {D~Grundler}(2014)}]{Krawczyk_2014}%
  \BibitemOpen
  \bibfield  {author} {\bibinfo {author} {\bibfnamefont {M.}~\bibnamefont
  {Krawczyk}}\ and\ \bibinfo {author} {\bibfnamefont {D.}~\bibnamefont
  {D~Grundler}},\ }\bibfield  {title} {\bibinfo {title} {Review and prospects
  of magnonic crystals and devices with reprogrammable band structure},\ }\href
  {https://doi.org/10.1088/0953-8984/26/12/123202} {\bibfield  {journal}
  {\bibinfo  {journal} {J. Phys.: Condens. Matter}\ }\textbf {\bibinfo {volume}
  {26}},\ \bibinfo {pages} {123202} (\bibinfo {year} {2014})}\BibitemShut
  {NoStop}%
\bibitem [{\citenamefont {Chumak}\ \emph {et~al.}(2015)\citenamefont {Chumak},
  \citenamefont {Vasyuchka}, \citenamefont {Serga},\ and\ \citenamefont
  {Hillebrands}}]{Chumak_2015}%
  \BibitemOpen
  \bibfield  {author} {\bibinfo {author} {\bibfnamefont {A.~V.}\ \bibnamefont
  {Chumak}}, \bibinfo {author} {\bibfnamefont {V.~I.}\ \bibnamefont
  {Vasyuchka}}, \bibinfo {author} {\bibfnamefont {A.~A.}\ \bibnamefont
  {Serga}},\ and\ \bibinfo {author} {\bibfnamefont {B.}~\bibnamefont
  {Hillebrands}},\ }\bibfield  {title} {\bibinfo {title} {Magnon spintronics},\
  }\href {https://doi.org/https://doi.org/10.1038/nphys3347} {\bibfield
  {journal} {\bibinfo  {journal} {Nat. Phys.}\ }\textbf {\bibinfo {volume}
  {11}},\ \bibinfo {pages} {453} (\bibinfo {year} {2015})}\BibitemShut
  {NoStop}%
\bibitem [{\citenamefont {Nikitov}\ \emph {et~al.}(2001)\citenamefont
  {Nikitov}, \citenamefont {Tailhades},\ and\ \citenamefont
  {Tsai}}]{Nikitov_2001}%
  \BibitemOpen
  \bibfield  {author} {\bibinfo {author} {\bibfnamefont {S.~A.}\ \bibnamefont
  {Nikitov}}, \bibinfo {author} {\bibfnamefont {P.}~\bibnamefont {Tailhades}},\
  and\ \bibinfo {author} {\bibfnamefont {C.~S.}\ \bibnamefont {Tsai}},\
  }\bibfield  {title} {\bibinfo {title} {Spin waves in periodic magnetic
  structures -- magnonic crystals},\ }\href
  {https://doi.org/https://doi.org/10.1016/S0304-8853(01)00470-X} {\bibfield
  {journal} {\bibinfo  {journal} {J. Magn. Magn. Mater.}\ }\textbf {\bibinfo
  {volume} {236}},\ \bibinfo {pages} {320 } (\bibinfo {year}
  {2001})}\BibitemShut {NoStop}%
\bibitem [{\citenamefont {Kostylev}\ \emph {et~al.}(2004)\citenamefont
  {Kostylev}, \citenamefont {Stashkevich},\ and\ \citenamefont
  {Sergeeva}}]{Kostylev_2004}%
  \BibitemOpen
  \bibfield  {author} {\bibinfo {author} {\bibfnamefont {M.~P.}\ \bibnamefont
  {Kostylev}}, \bibinfo {author} {\bibfnamefont {A.~A.}\ \bibnamefont
  {Stashkevich}},\ and\ \bibinfo {author} {\bibfnamefont {N.~A.}\ \bibnamefont
  {Sergeeva}},\ }\bibfield  {title} {\bibinfo {title} {Collective magnetostatic
  modes on a one-dimensional array of ferromagnetic stripes},\ }\href
  {https://doi.org/10.1103/PhysRevB.69.064408} {\bibfield  {journal} {\bibinfo
  {journal} {Phys. Rev. B}\ }\textbf {\bibinfo {volume} {69}},\ \bibinfo
  {pages} {064408} (\bibinfo {year} {2004})}\BibitemShut {NoStop}%
\bibitem [{\citenamefont {Guslienko}\ and\ \citenamefont
  {Slavin}(2005)}]{Guslienko_2005}%
  \BibitemOpen
  \bibfield  {author} {\bibinfo {author} {\bibfnamefont {K.~Y.}\ \bibnamefont
  {Guslienko}}\ and\ \bibinfo {author} {\bibfnamefont {A.~N.}\ \bibnamefont
  {Slavin}},\ }\bibfield  {title} {\bibinfo {title} {Boundary conditions for
  magnetization in magnetic nanoelements},\ }\href
  {https://doi.org/10.1103/PhysRevB.72.014463} {\bibfield  {journal} {\bibinfo
  {journal} {Phys. Rev. B}\ }\textbf {\bibinfo {volume} {72}},\ \bibinfo
  {pages} {014463} (\bibinfo {year} {2005})}\BibitemShut {NoStop}%
\bibitem [{\citenamefont {Gubbiotti}\ \emph {et~al.}(2005)\citenamefont
  {Gubbiotti}, \citenamefont {Tacchi}, \citenamefont {Carlotti}, \citenamefont
  {Vavassori}, \citenamefont {Singh}, \citenamefont {Goolaup}, \citenamefont
  {Adeyeye}, \citenamefont {Stashkevich},\ and\ \citenamefont
  {Kostylev}}]{Gubbiotti_2005}%
  \BibitemOpen
  \bibfield  {author} {\bibinfo {author} {\bibfnamefont {G.}~\bibnamefont
  {Gubbiotti}}, \bibinfo {author} {\bibfnamefont {S.}~\bibnamefont {Tacchi}},
  \bibinfo {author} {\bibfnamefont {G.}~\bibnamefont {Carlotti}}, \bibinfo
  {author} {\bibfnamefont {P.}~\bibnamefont {Vavassori}}, \bibinfo {author}
  {\bibfnamefont {N.}~\bibnamefont {Singh}}, \bibinfo {author} {\bibfnamefont
  {S.}~\bibnamefont {Goolaup}}, \bibinfo {author} {\bibfnamefont {A.~O.}\
  \bibnamefont {Adeyeye}}, \bibinfo {author} {\bibfnamefont {A.}~\bibnamefont
  {Stashkevich}},\ and\ \bibinfo {author} {\bibfnamefont {M.}~\bibnamefont
  {Kostylev}},\ }\bibfield  {title} {\bibinfo {title} {Magnetostatic
  interaction in arrays of nanometric permalloy wires: A magneto-optic kerr
  effect and a brillouin light scattering study},\ }\href
  {https://doi.org/10.1103/PhysRevB.72.224413} {\bibfield  {journal} {\bibinfo
  {journal} {Phys. Rev. B}\ }\textbf {\bibinfo {volume} {72}},\ \bibinfo
  {pages} {224413} (\bibinfo {year} {2005})}\BibitemShut {NoStop}%
\bibitem [{\citenamefont {Gubbiotti}\ \emph {et~al.}(2007)\citenamefont
  {Gubbiotti}, \citenamefont {Tacchi}, \citenamefont {Carlotti}, \citenamefont
  {Singh}, \citenamefont {Goolaup}, \citenamefont {Adeyeye},\ and\
  \citenamefont {Kostylev}}]{Gubbiotti_2007}%
  \BibitemOpen
  \bibfield  {author} {\bibinfo {author} {\bibfnamefont {G.}~\bibnamefont
  {Gubbiotti}}, \bibinfo {author} {\bibfnamefont {S.}~\bibnamefont {Tacchi}},
  \bibinfo {author} {\bibfnamefont {G.}~\bibnamefont {Carlotti}}, \bibinfo
  {author} {\bibfnamefont {N.}~\bibnamefont {Singh}}, \bibinfo {author}
  {\bibfnamefont {S.}~\bibnamefont {Goolaup}}, \bibinfo {author} {\bibfnamefont
  {A.~O.}\ \bibnamefont {Adeyeye}},\ and\ \bibinfo {author} {\bibfnamefont
  {M.}~\bibnamefont {Kostylev}},\ }\bibfield  {title} {\bibinfo {title}
  {Collective spin modes in monodimensional magnonic crystals consisting of
  dipolarly coupled nanowires},\ }\href
  {https://doi.org/https://doi.org/10.1063/1.2709909} {\bibfield  {journal}
  {\bibinfo  {journal} {Appl. Phys. Lett.}\ }\textbf {\bibinfo {volume} {90}},\
  \bibinfo {pages} {092503} (\bibinfo {year} {2007})}\BibitemShut {NoStop}%
\bibitem [{\citenamefont {Kostylev}\ \emph {et~al.}(2008)\citenamefont
  {Kostylev}, \citenamefont {Schrader}, \citenamefont {Stamps}, \citenamefont
  {Gubbiotti}, \citenamefont {Carlotti}, \citenamefont {Adeyeye}, \citenamefont
  {Goolaup},\ and\ \citenamefont {Singh}}]{Kostylev_2008}%
  \BibitemOpen
  \bibfield  {author} {\bibinfo {author} {\bibfnamefont {M.}~\bibnamefont
  {Kostylev}}, \bibinfo {author} {\bibfnamefont {P.}~\bibnamefont {Schrader}},
  \bibinfo {author} {\bibfnamefont {R.~L.}\ \bibnamefont {Stamps}}, \bibinfo
  {author} {\bibfnamefont {G.}~\bibnamefont {Gubbiotti}}, \bibinfo {author}
  {\bibfnamefont {G.}~\bibnamefont {Carlotti}}, \bibinfo {author}
  {\bibfnamefont {A.~O.}\ \bibnamefont {Adeyeye}}, \bibinfo {author}
  {\bibfnamefont {S.}~\bibnamefont {Goolaup}},\ and\ \bibinfo {author}
  {\bibfnamefont {N.}~\bibnamefont {Singh}},\ }\bibfield  {title} {\bibinfo
  {title} {Partial frequency band gap in one-dimensional magnonic crystals},\
  }\href {https://doi.org/10.1063/1.2904697} {\bibfield  {journal} {\bibinfo
  {journal} {Appl. Phys. Lett.}\ }\textbf {\bibinfo {volume} {92}},\ \bibinfo
  {pages} {132504} (\bibinfo {year} {2008})}\BibitemShut {NoStop}%
\bibitem [{\citenamefont {Polushkin}(2008)}]{Polushkin_2008}%
  \BibitemOpen
  \bibfield  {author} {\bibinfo {author} {\bibfnamefont {N.~I.}\ \bibnamefont
  {Polushkin}},\ }\bibfield  {title} {\bibinfo {title} {Excitation of coupled
  oscillations in lateral ferromagnetic heterostructures},\ }\href
  {https://doi.org/10.1103/PhysRevB.77.180401} {\bibfield  {journal} {\bibinfo
  {journal} {Phys. Rev. B}\ }\textbf {\bibinfo {volume} {77}},\ \bibinfo
  {pages} {180401(R)} (\bibinfo {year} {2008})}\BibitemShut {NoStop}%
\bibitem [{\citenamefont {Mruczkiewicz}\ \emph
  {et~al.}(2013{\natexlab{a}})\citenamefont {Mruczkiewicz}, \citenamefont
  {Krawczyk}, \citenamefont {Sakharov}, \citenamefont {Khivintsev},
  \citenamefont {Filimonov},\ and\ \citenamefont
  {Nikitov}}]{Mruczkiewicz_Comsol}%
  \BibitemOpen
  \bibfield  {author} {\bibinfo {author} {\bibfnamefont {M.}~\bibnamefont
  {Mruczkiewicz}}, \bibinfo {author} {\bibfnamefont {M.}~\bibnamefont
  {Krawczyk}}, \bibinfo {author} {\bibfnamefont {V.~K.}\ \bibnamefont
  {Sakharov}}, \bibinfo {author} {\bibfnamefont {Y.~V.}\ \bibnamefont
  {Khivintsev}}, \bibinfo {author} {\bibfnamefont {Y.~A.}\ \bibnamefont
  {Filimonov}},\ and\ \bibinfo {author} {\bibfnamefont {S.~A.}\ \bibnamefont
  {Nikitov}},\ }\bibfield  {title} {\bibinfo {title} {Standing spin waves in
  magnonic crystals},\ }\href {https://doi.org/10.1063/1.4793085} {\bibfield
  {journal} {\bibinfo  {journal} {J. Appl. Phys.}\ }\textbf {\bibinfo {volume}
  {113}},\ \bibinfo {pages} {093908} (\bibinfo {year}
  {2013}{\natexlab{a}})}\BibitemShut {NoStop}%
\bibitem [{\citenamefont {Mruczkiewicz}\ \emph
  {et~al.}(2013{\natexlab{b}})\citenamefont {Mruczkiewicz}, \citenamefont
  {Krawczyk}, \citenamefont {Gubbiotti}, \citenamefont {Tacchi}, \citenamefont
  {Filimonov}, \citenamefont {Kalyabin}, \citenamefont {Lisenkov},\ and\
  \citenamefont {Nikitov}}]{Mruczkiewicz_2013}%
  \BibitemOpen
  \bibfield  {author} {\bibinfo {author} {\bibfnamefont {M.}~\bibnamefont
  {Mruczkiewicz}}, \bibinfo {author} {\bibfnamefont {M.}~\bibnamefont
  {Krawczyk}}, \bibinfo {author} {\bibfnamefont {G.}~\bibnamefont {Gubbiotti}},
  \bibinfo {author} {\bibfnamefont {S.}~\bibnamefont {Tacchi}}, \bibinfo
  {author} {\bibfnamefont {Y.~A.}\ \bibnamefont {Filimonov}}, \bibinfo {author}
  {\bibfnamefont {D.~V.}\ \bibnamefont {Kalyabin}}, \bibinfo {author}
  {\bibfnamefont {I.~V.}\ \bibnamefont {Lisenkov}},\ and\ \bibinfo {author}
  {\bibfnamefont {S.~A.}\ \bibnamefont {Nikitov}},\ }\bibfield  {title}
  {\bibinfo {title} {Nonreciprocity of spin waves in metallized magnonic
  crystal},\ }\href {https://doi.org/10.1088/1367-2630/15/11/113023} {\bibfield
   {journal} {\bibinfo  {journal} {New J. Phys.}\ }\textbf {\bibinfo {volume}
  {15}},\ \bibinfo {pages} {113023} (\bibinfo {year}
  {2013}{\natexlab{b}})}\BibitemShut {NoStop}%
\bibitem [{\citenamefont {Lisenkov}\ \emph {et~al.}(2015)\citenamefont
  {Lisenkov}, \citenamefont {Kalyabin}, \citenamefont {Osokin}, \citenamefont
  {Klos}, \citenamefont {Krawczyk},\ and\ \citenamefont
  {Nikitov}}]{Lisenkov_2015}%
  \BibitemOpen
  \bibfield  {author} {\bibinfo {author} {\bibfnamefont {I.}~\bibnamefont
  {Lisenkov}}, \bibinfo {author} {\bibfnamefont {D.}~\bibnamefont {Kalyabin}},
  \bibinfo {author} {\bibfnamefont {S.}~\bibnamefont {Osokin}}, \bibinfo
  {author} {\bibfnamefont {J.~W.}\ \bibnamefont {Klos}}, \bibinfo {author}
  {\bibfnamefont {M.}~\bibnamefont {Krawczyk}},\ and\ \bibinfo {author}
  {\bibfnamefont {S.}~\bibnamefont {Nikitov}},\ }\bibfield  {title} {\bibinfo
  {title} {Nonreciprocity of edge modes in 1d magnonic crystal},\ }\href
  {https://doi.org/https://doi.org/10.1016/j.jmmm.2014.10.073} {\bibfield
  {journal} {\bibinfo  {journal} {J. Magn. Magn. Mater.}\ }\textbf {\bibinfo
  {volume} {378}},\ \bibinfo {pages} {313} (\bibinfo {year}
  {2015})}\BibitemShut {NoStop}%
\bibitem [{\citenamefont {Rych{\l}y}\ \emph {et~al.}(2015)\citenamefont
  {Rych{\l}y}, \citenamefont {Gruszecki}, \citenamefont {Mruczkiewicz},
  \citenamefont {K{\l}os}, \citenamefont {Mamica},\ and\ \citenamefont
  {Krawczyk}}]{Rychly_2015}%
  \BibitemOpen
  \bibfield  {author} {\bibinfo {author} {\bibfnamefont {J.}~\bibnamefont
  {Rych{\l}y}}, \bibinfo {author} {\bibfnamefont {P.}~\bibnamefont
  {Gruszecki}}, \bibinfo {author} {\bibfnamefont {M.}~\bibnamefont
  {Mruczkiewicz}}, \bibinfo {author} {\bibfnamefont {J.~W.}\ \bibnamefont
  {K{\l}os}}, \bibinfo {author} {\bibfnamefont {S.}~\bibnamefont {Mamica}},\
  and\ \bibinfo {author} {\bibfnamefont {M.}~\bibnamefont {Krawczyk}},\
  }\bibfield  {title} {\bibinfo {title} {Magnonic crystals -- prospective
  structures for shaping spin waves in nanoscale},\ }\href
  {https://doi.org/10.1063/1.4932348} {\bibfield  {journal} {\bibinfo
  {journal} {Low Temperature Physics}\ }\textbf {\bibinfo {volume} {41}},\
  \bibinfo {pages} {745} (\bibinfo {year} {2015})}\BibitemShut {NoStop}%
\bibitem [{\citenamefont {Chumak}\ \emph {et~al.}(2008)\citenamefont {Chumak},
  \citenamefont {Serga}, \citenamefont {Hillebrands},\ and\ \citenamefont
  {Kostylev}}]{Chumak_2008}%
  \BibitemOpen
  \bibfield  {author} {\bibinfo {author} {\bibfnamefont {A.~V.}\ \bibnamefont
  {Chumak}}, \bibinfo {author} {\bibfnamefont {A.~A.}\ \bibnamefont {Serga}},
  \bibinfo {author} {\bibfnamefont {B.}~\bibnamefont {Hillebrands}},\ and\
  \bibinfo {author} {\bibfnamefont {M.~P.}\ \bibnamefont {Kostylev}},\
  }\bibfield  {title} {\bibinfo {title} {Scattering of backward spin waves in a
  one-dimensional magnonic crystal},\ }\href
  {https://doi.org/10.1063/1.2963027} {\bibfield  {journal} {\bibinfo
  {journal} {Appl. Phys. Lett.}\ }\textbf {\bibinfo {volume} {93}},\ \bibinfo
  {pages} {022508} (\bibinfo {year} {2008})}\BibitemShut {NoStop}%
\bibitem [{\citenamefont {Chumak}\ \emph
  {et~al.}(2009{\natexlab{a}})\citenamefont {Chumak}, \citenamefont {Serga},
  \citenamefont {Wolff}, \citenamefont {Hillebrands},\ and\ \citenamefont
  {Kostylev}}]{Chumak_2009}%
  \BibitemOpen
  \bibfield  {author} {\bibinfo {author} {\bibfnamefont {A.~V.}\ \bibnamefont
  {Chumak}}, \bibinfo {author} {\bibfnamefont {A.~A.}\ \bibnamefont {Serga}},
  \bibinfo {author} {\bibfnamefont {S.}~\bibnamefont {Wolff}}, \bibinfo
  {author} {\bibfnamefont {B.}~\bibnamefont {Hillebrands}},\ and\ \bibinfo
  {author} {\bibfnamefont {M.~P.}\ \bibnamefont {Kostylev}},\ }\bibfield
  {title} {\bibinfo {title} {Design and optimization of one-dimensional
  ferrite-film based magnonic crystals},\ }\href
  {https://doi.org/10.1063/1.3098258} {\bibfield  {journal} {\bibinfo
  {journal} {J. Appl. Phys.}\ }\textbf {\bibinfo {volume} {105}},\ \bibinfo
  {pages} {083906} (\bibinfo {year} {2009}{\natexlab{a}})}\BibitemShut
  {NoStop}%
\bibitem [{\citenamefont {Chumak}\ \emph
  {et~al.}(2009{\natexlab{b}})\citenamefont {Chumak}, \citenamefont {Serga},
  \citenamefont {Wolff}, \citenamefont {Hillebrands},\ and\ \citenamefont
  {Kostylev}}]{Chumak_2009APL}%
  \BibitemOpen
  \bibfield  {author} {\bibinfo {author} {\bibfnamefont {A.~V.}\ \bibnamefont
  {Chumak}}, \bibinfo {author} {\bibfnamefont {A.~A.}\ \bibnamefont {Serga}},
  \bibinfo {author} {\bibfnamefont {S.}~\bibnamefont {Wolff}}, \bibinfo
  {author} {\bibfnamefont {B.}~\bibnamefont {Hillebrands}},\ and\ \bibinfo
  {author} {\bibfnamefont {M.~P.}\ \bibnamefont {Kostylev}},\ }\bibfield
  {title} {\bibinfo {title} {Scattering of surface and volume spin waves in a
  magnonic crystal},\ }\href {https://doi.org/10.1063/1.3127227} {\bibfield
  {journal} {\bibinfo  {journal} {Appl. Phys. Lett.}\ }\textbf {\bibinfo
  {volume} {94}},\ \bibinfo {pages} {172511} (\bibinfo {year}
  {2009}{\natexlab{b}})}\BibitemShut {NoStop}%
\bibitem [{\citenamefont {Serga}\ \emph {et~al.}(2010)\citenamefont {Serga},
  \citenamefont {Chumak},\ and\ \citenamefont {Hillebrands}}]{Serga_2010}%
  \BibitemOpen
  \bibfield  {author} {\bibinfo {author} {\bibfnamefont {A.~A.}\ \bibnamefont
  {Serga}}, \bibinfo {author} {\bibfnamefont {A.~V.}\ \bibnamefont {Chumak}},\
  and\ \bibinfo {author} {\bibfnamefont {B.}~\bibnamefont {Hillebrands}},\
  }\bibfield  {title} {\bibinfo {title} {{YIG} magnonics},\ }\href
  {https://doi.org/10.1088/0022-3727/43/26/264002} {\bibfield  {journal}
  {\bibinfo  {journal} {J. Phys. D: Appl. Phys.}\ }\textbf {\bibinfo {volume}
  {43}},\ \bibinfo {pages} {264002} (\bibinfo {year} {2010})}\BibitemShut
  {NoStop}%
\bibitem [{\citenamefont {Landeros}\ and\ \citenamefont
  {Mills}(2012)}]{Landeros_2012}%
  \BibitemOpen
  \bibfield  {author} {\bibinfo {author} {\bibfnamefont {P.}~\bibnamefont
  {Landeros}}\ and\ \bibinfo {author} {\bibfnamefont {D.~L.}\ \bibnamefont
  {Mills}},\ }\bibfield  {title} {\bibinfo {title} {Spin waves in periodically
  perturbed films},\ }\href {https://doi.org/10.1103/PhysRevB.85.054424}
  {\bibfield  {journal} {\bibinfo  {journal} {Phys. Rev. B}\ }\textbf {\bibinfo
  {volume} {85}},\ \bibinfo {pages} {054424} (\bibinfo {year}
  {2012})}\BibitemShut {NoStop}%
\bibitem [{\citenamefont {Kakazei}\ \emph {et~al.}(2014)\citenamefont
  {Kakazei}, \citenamefont {Liu}, \citenamefont {Ding},\ and\ \citenamefont
  {Adeyeye}}]{Kakazei_2014}%
  \BibitemOpen
  \bibfield  {author} {\bibinfo {author} {\bibfnamefont {G.~N.}\ \bibnamefont
  {Kakazei}}, \bibinfo {author} {\bibfnamefont {X.~M.}\ \bibnamefont {Liu}},
  \bibinfo {author} {\bibfnamefont {J.}~\bibnamefont {Ding}},\ and\ \bibinfo
  {author} {\bibfnamefont {A.~O.}\ \bibnamefont {Adeyeye}},\ }\bibfield
  {title} {\bibinfo {title} {Ni80fe20 film with periodically modulated
  thickness as a reconfigurable one-dimensional magnonic crystal},\ }\href
  {https://doi.org/10.1063/1.4863508} {\bibfield  {journal} {\bibinfo
  {journal} {Appl. Phys. Lett.}\ }\textbf {\bibinfo {volume} {104}},\ \bibinfo
  {pages} {042403} (\bibinfo {year} {2014})}\BibitemShut {NoStop}%
\bibitem [{\citenamefont {Aranda}\ \emph {et~al.}(2014)\citenamefont {Aranda},
  \citenamefont {Kakazei}, \citenamefont {Gonz\'{a}lez},\ and\ \citenamefont
  {Guslienko}}]{Aranda_2014}%
  \BibitemOpen
  \bibfield  {author} {\bibinfo {author} {\bibfnamefont {G.~R.}\ \bibnamefont
  {Aranda}}, \bibinfo {author} {\bibfnamefont {G.~N.}\ \bibnamefont {Kakazei}},
  \bibinfo {author} {\bibfnamefont {J.}~\bibnamefont {Gonz\'{a}lez}},\ and\
  \bibinfo {author} {\bibfnamefont {K.~Y.}\ \bibnamefont {Guslienko}},\
  }\bibfield  {title} {\bibinfo {title} {Ferromagnetic resonance micromagnetic
  studies in patterned permalloy thin films and stripes},\ }\href
  {https://doi.org/10.1063/1.4894164} {\bibfield  {journal} {\bibinfo
  {journal} {J. Appl. Phys.}\ }\textbf {\bibinfo {volume} {116}},\ \bibinfo
  {pages} {093908} (\bibinfo {year} {2014})}\BibitemShut {NoStop}%
\bibitem [{\citenamefont {Bessonov}\ \emph {et~al.}(2015)\citenamefont
  {Bessonov}, \citenamefont {Mruczkiewicz}, \citenamefont {Gieniusz},
  \citenamefont {Guzowska}, \citenamefont {Maziewski}, \citenamefont
  {Stognij},\ and\ \citenamefont {Krawczyk}}]{Bessonov_2015}%
  \BibitemOpen
  \bibfield  {author} {\bibinfo {author} {\bibfnamefont {V.~D.}\ \bibnamefont
  {Bessonov}}, \bibinfo {author} {\bibfnamefont {M.}~\bibnamefont
  {Mruczkiewicz}}, \bibinfo {author} {\bibfnamefont {R.}~\bibnamefont
  {Gieniusz}}, \bibinfo {author} {\bibfnamefont {U.}~\bibnamefont {Guzowska}},
  \bibinfo {author} {\bibfnamefont {A.}~\bibnamefont {Maziewski}}, \bibinfo
  {author} {\bibfnamefont {A.~I.}\ \bibnamefont {Stognij}},\ and\ \bibinfo
  {author} {\bibfnamefont {M.}~\bibnamefont {Krawczyk}},\ }\bibfield  {title}
  {\bibinfo {title} {Magnonic band gaps in yig-based one-dimensional magnonic
  crystals: An array of grooves versus an array of metallic stripes},\ }\href
  {https://doi.org/10.1103/PhysRevB.91.104421} {\bibfield  {journal} {\bibinfo
  {journal} {Phys. Rev. B}\ }\textbf {\bibinfo {volume} {91}},\ \bibinfo
  {pages} {104421} (\bibinfo {year} {2015})}\BibitemShut {NoStop}%
\bibitem [{\citenamefont {Langer}\ \emph {et~al.}(2017)\citenamefont {Langer},
  \citenamefont {R\"oder}, \citenamefont {Gallardo}, \citenamefont {Schneider},
  \citenamefont {Stienen}, \citenamefont {Gatel}, \citenamefont {H\"ubner},
  \citenamefont {Bischoff}, \citenamefont {Lenz}, \citenamefont {Lindner},
  \citenamefont {Landeros},\ and\ \citenamefont {Fassbender}}]{Langer_2017}%
  \BibitemOpen
  \bibfield  {author} {\bibinfo {author} {\bibfnamefont {M.}~\bibnamefont
  {Langer}}, \bibinfo {author} {\bibfnamefont {F.}~\bibnamefont {R\"oder}},
  \bibinfo {author} {\bibfnamefont {R.~A.}\ \bibnamefont {Gallardo}}, \bibinfo
  {author} {\bibfnamefont {T.}~\bibnamefont {Schneider}}, \bibinfo {author}
  {\bibfnamefont {S.}~\bibnamefont {Stienen}}, \bibinfo {author} {\bibfnamefont
  {C.}~\bibnamefont {Gatel}}, \bibinfo {author} {\bibfnamefont
  {R.}~\bibnamefont {H\"ubner}}, \bibinfo {author} {\bibfnamefont
  {L.}~\bibnamefont {Bischoff}}, \bibinfo {author} {\bibfnamefont
  {K.}~\bibnamefont {Lenz}}, \bibinfo {author} {\bibfnamefont {J.}~\bibnamefont
  {Lindner}}, \bibinfo {author} {\bibfnamefont {P.}~\bibnamefont {Landeros}},\
  and\ \bibinfo {author} {\bibfnamefont {J.}~\bibnamefont {Fassbender}},\
  }\bibfield  {title} {\bibinfo {title} {Role of internal demagnetizing field
  for the dynamics of a surface-modulated magnonic crystal},\ }\href
  {https://doi.org/10.1103/PhysRevB.95.184405} {\bibfield  {journal} {\bibinfo
  {journal} {Phys. Rev. B}\ }\textbf {\bibinfo {volume} {95}},\ \bibinfo
  {pages} {184405} (\bibinfo {year} {2017})}\BibitemShut {NoStop}%
\bibitem [{\citenamefont {Gallardo}\ \emph {et~al.}(2018)\citenamefont
  {Gallardo}, \citenamefont {Schneider}, \citenamefont {Rold\'an-Molina},
  \citenamefont {Langer}, \citenamefont {Fassbender}, \citenamefont {Lenz},
  \citenamefont {Lindner},\ and\ \citenamefont {Landeros}}]{Gallardo_2018}%
  \BibitemOpen
  \bibfield  {author} {\bibinfo {author} {\bibfnamefont {R.~A.}\ \bibnamefont
  {Gallardo}}, \bibinfo {author} {\bibfnamefont {T.}~\bibnamefont {Schneider}},
  \bibinfo {author} {\bibfnamefont {A.}~\bibnamefont {Rold\'an-Molina}},
  \bibinfo {author} {\bibfnamefont {M.}~\bibnamefont {Langer}}, \bibinfo
  {author} {\bibfnamefont {J.}~\bibnamefont {Fassbender}}, \bibinfo {author}
  {\bibfnamefont {K.}~\bibnamefont {Lenz}}, \bibinfo {author} {\bibfnamefont
  {J.}~\bibnamefont {Lindner}},\ and\ \bibinfo {author} {\bibfnamefont
  {P.}~\bibnamefont {Landeros}},\ }\bibfield  {title} {\bibinfo {title}
  {Dipolar interaction induced band gaps and flat modes in surface-modulated
  magnonic crystals},\ }\href
  {https://doi.org/https://doi.org/10.1103/PhysRevB.97.144405} {\bibfield
  {journal} {\bibinfo  {journal} {Phys. Rev. B}\ }\textbf {\bibinfo {volume}
  {97}},\ \bibinfo {pages} {144405} (\bibinfo {year} {2018})}\BibitemShut
  {NoStop}%
\bibitem [{\citenamefont {Langer}\ \emph {et~al.}(2019)\citenamefont {Langer},
  \citenamefont {Gallardo}, \citenamefont {Schneider}, \citenamefont {Stienen},
  \citenamefont {Rold{\'{a}}n-Molina}, \citenamefont {Yuan}, \citenamefont
  {Lenz}, \citenamefont {Lindner}, \citenamefont {Landeros},\ and\
  \citenamefont {Fassbender}}]{Langer_2019}%
  \BibitemOpen
  \bibfield  {author} {\bibinfo {author} {\bibfnamefont {M.}~\bibnamefont
  {Langer}}, \bibinfo {author} {\bibfnamefont {R.~A.}\ \bibnamefont
  {Gallardo}}, \bibinfo {author} {\bibfnamefont {T.}~\bibnamefont {Schneider}},
  \bibinfo {author} {\bibfnamefont {S.}~\bibnamefont {Stienen}}, \bibinfo
  {author} {\bibfnamefont {A.}~\bibnamefont {Rold{\'{a}}n-Molina}}, \bibinfo
  {author} {\bibfnamefont {Y.}~\bibnamefont {Yuan}}, \bibinfo {author}
  {\bibfnamefont {K.}~\bibnamefont {Lenz}}, \bibinfo {author} {\bibfnamefont
  {J.}~\bibnamefont {Lindner}}, \bibinfo {author} {\bibfnamefont
  {P.}~\bibnamefont {Landeros}},\ and\ \bibinfo {author} {\bibfnamefont
  {J.}~\bibnamefont {Fassbender}},\ }\bibfield  {title} {\bibinfo {title}
  {Spin-wave modes in transition from a thin film to a full magnonic crystal},\
  }\href {https://doi.org/10.1103/physrevb.99.024426} {\bibfield  {journal}
  {\bibinfo  {journal} {Phys. Rev. B}\ }\textbf {\bibinfo {volume} {99}},\
  \bibinfo {pages} {024426} (\bibinfo {year} {2019})}\BibitemShut {NoStop}%
\bibitem [{\citenamefont {Gallardo}\ \emph {et~al.}(2019)\citenamefont
  {Gallardo}, \citenamefont {Cort\'es-Ortu\~no}, \citenamefont {Schneider},
  \citenamefont {Rold\'an-Molina}, \citenamefont {Ma}, \citenamefont
  {Troncoso}, \citenamefont {Lenz}, \citenamefont {Fangohr}, \citenamefont
  {Lindner},\ and\ \citenamefont {Landeros}}]{Gallardo_2019}%
  \BibitemOpen
  \bibfield  {author} {\bibinfo {author} {\bibfnamefont {R.~A.}\ \bibnamefont
  {Gallardo}}, \bibinfo {author} {\bibfnamefont {D.}~\bibnamefont
  {Cort\'es-Ortu\~no}}, \bibinfo {author} {\bibfnamefont {T.}~\bibnamefont
  {Schneider}}, \bibinfo {author} {\bibfnamefont {A.}~\bibnamefont
  {Rold\'an-Molina}}, \bibinfo {author} {\bibfnamefont {F.}~\bibnamefont {Ma}},
  \bibinfo {author} {\bibfnamefont {R.~E.}\ \bibnamefont {Troncoso}}, \bibinfo
  {author} {\bibfnamefont {K.}~\bibnamefont {Lenz}}, \bibinfo {author}
  {\bibfnamefont {H.}~\bibnamefont {Fangohr}}, \bibinfo {author} {\bibfnamefont
  {J.}~\bibnamefont {Lindner}},\ and\ \bibinfo {author} {\bibfnamefont
  {P.}~\bibnamefont {Landeros}},\ }\bibfield  {title} {\bibinfo {title} {Flat
  bands, indirect gaps, and unconventional spin-wave behavior induced by a
  periodic dzyaloshinskii-moriya interaction},\ }\href
  {https://doi.org/10.1103/PhysRevLett.122.067204} {\bibfield  {journal}
  {\bibinfo  {journal} {Phys. Rev. Lett.}\ }\textbf {\bibinfo {volume} {122}},\
  \bibinfo {pages} {067204} (\bibinfo {year} {2019})}\BibitemShut {NoStop}%
\bibitem [{\citenamefont {Chumak}\ \emph {et~al.}(2014)\citenamefont {Chumak},
  \citenamefont {Serga},\ and\ \citenamefont {Hillebrands}}]{Chumak_2014}%
  \BibitemOpen
  \bibfield  {author} {\bibinfo {author} {\bibfnamefont {A.~V.}\ \bibnamefont
  {Chumak}}, \bibinfo {author} {\bibfnamefont {A.~A.}\ \bibnamefont {Serga}},\
  and\ \bibinfo {author} {\bibfnamefont {B.}~\bibnamefont {Hillebrands}},\
  }\bibfield  {title} {\bibinfo {title} {Magnon transistor for all-magnon data
  processing},\ }\href {https://doi.org/https://doi.org/10.1038/ncomms5700}
  {\bibfield  {journal} {\bibinfo  {journal} {Nat. Comm.}\ }\textbf {\bibinfo
  {volume} {5}},\ \bibinfo {pages} {4700} (\bibinfo {year} {2014})}\BibitemShut
  {NoStop}%
\bibitem [{\citenamefont {Khitun}\ \emph {et~al.}(2010)\citenamefont {Khitun},
  \citenamefont {Bao},\ and\ \citenamefont {Wang}}]{Khitun_2010}%
  \BibitemOpen
  \bibfield  {author} {\bibinfo {author} {\bibfnamefont {A.}~\bibnamefont
  {Khitun}}, \bibinfo {author} {\bibfnamefont {M.}~\bibnamefont {Bao}},\ and\
  \bibinfo {author} {\bibfnamefont {K.~L.}\ \bibnamefont {Wang}},\ }\bibfield
  {title} {\bibinfo {title} {Magnonic logic circuits},\ }\href
  {https://doi.org/10.1088/0022-3727/43/26/264005} {\bibfield  {journal}
  {\bibinfo  {journal} {J. Phys. D: Appl. Phys.}\ }\textbf {\bibinfo {volume}
  {43}},\ \bibinfo {pages} {264005} (\bibinfo {year} {2010})}\BibitemShut
  {NoStop}%
\bibitem [{\citenamefont {Vogt}\ \emph {et~al.}(2014)\citenamefont {Vogt},
  \citenamefont {Fradin}, \citenamefont {Pearson}, \citenamefont {Sebastian},
  \citenamefont {Bader}, \citenamefont {Hillebrands},\ and\ \citenamefont
  {Hoffmann}}]{Vogt_2014}%
  \BibitemOpen
  \bibfield  {author} {\bibinfo {author} {\bibfnamefont {K.}~\bibnamefont
  {Vogt}}, \bibinfo {author} {\bibfnamefont {F.~Y.}\ \bibnamefont {Fradin}},
  \bibinfo {author} {\bibfnamefont {J.~E.}\ \bibnamefont {Pearson}}, \bibinfo
  {author} {\bibfnamefont {T.}~\bibnamefont {Sebastian}}, \bibinfo {author}
  {\bibfnamefont {S.~D.}\ \bibnamefont {Bader}}, \bibinfo {author}
  {\bibfnamefont {B.}~\bibnamefont {Hillebrands}},\ and\ \bibinfo {author}
  {\bibfnamefont {H.}~\bibnamefont {Hoffmann}, \bibfnamefont
  {A.~Schultheiss}},\ }\bibfield  {title} {\bibinfo {title} {Realization of a
  spin-wave multiplexer},\ }\href
  {https://doi.org/https://doi.org/10.1038/ncomms4727} {\bibfield  {journal}
  {\bibinfo  {journal} {Nat. Comm.}\ }\textbf {\bibinfo {volume} {5}},\
  \bibinfo {pages} {3727} (\bibinfo {year} {2014})}\BibitemShut {NoStop}%
\bibitem [{\citenamefont {Balinskiy}\ \emph {et~al.}(2018)\citenamefont
  {Balinskiy}, \citenamefont {Chiang},\ and\ \citenamefont
  {A.}}]{Balinskiy_2018}%
  \BibitemOpen
  \bibfield  {author} {\bibinfo {author} {\bibfnamefont {M.}~\bibnamefont
  {Balinskiy}}, \bibinfo {author} {\bibfnamefont {H.}~\bibnamefont {Chiang}},\
  and\ \bibinfo {author} {\bibfnamefont {K.}~\bibnamefont {A.}},\ }\bibfield
  {title} {\bibinfo {title} {Realization of spin wave switch for data
  processing},\ }\href {https://doi.org/https://doi.org/10.1063/1.5007164}
  {\bibfield  {journal} {\bibinfo  {journal} {AIP Adv.}\ }\textbf {\bibinfo
  {volume} {8}},\ \bibinfo {pages} {056628} (\bibinfo {year}
  {2018})}\BibitemShut {NoStop}%
\bibitem [{\citenamefont {Kim}\ \emph {et~al.}(2009)\citenamefont {Kim},
  \citenamefont {Lee},\ and\ \citenamefont {Han}}]{Kim_2009}%
  \BibitemOpen
  \bibfield  {author} {\bibinfo {author} {\bibfnamefont {S.-K.}\ \bibnamefont
  {Kim}}, \bibinfo {author} {\bibfnamefont {K.-S.}\ \bibnamefont {Lee}},\ and\
  \bibinfo {author} {\bibfnamefont {D.-S.}\ \bibnamefont {Han}},\ }\bibfield
  {title} {\bibinfo {title} {A gigahertz-range spin-wave filter composed of
  width-modulated nanostrip magnonic-crystal waveguides},\ }\href
  {https://doi.org/https://doi.org/10.1063/1.3186782} {\bibfield  {journal}
  {\bibinfo  {journal} {Appl. Phys. Lett}\ }\textbf {\bibinfo {volume} {95}},\
  \bibinfo {pages} {082507} (\bibinfo {year} {2009})}\BibitemShut {NoStop}%
\bibitem [{\citenamefont {Yu}\ \emph {et~al.}(2013)\citenamefont {Yu},
  \citenamefont {Duerr}, \citenamefont {Huber}, \citenamefont {Bahr},
  \citenamefont {Schwarze}, \citenamefont {Brandl},\ and\ \citenamefont
  {Grundler}}]{Yu_2013}%
  \BibitemOpen
  \bibfield  {author} {\bibinfo {author} {\bibfnamefont {H.}~\bibnamefont
  {Yu}}, \bibinfo {author} {\bibfnamefont {G.}~\bibnamefont {Duerr}}, \bibinfo
  {author} {\bibfnamefont {R.}~\bibnamefont {Huber}}, \bibinfo {author}
  {\bibfnamefont {M.}~\bibnamefont {Bahr}}, \bibinfo {author} {\bibfnamefont
  {T.}~\bibnamefont {Schwarze}}, \bibinfo {author} {\bibfnamefont
  {F.}~\bibnamefont {Brandl}},\ and\ \bibinfo {author} {\bibfnamefont
  {D.}~\bibnamefont {Grundler}},\ }\bibfield  {title} {\bibinfo {title}
  {Omnidirectional spin-wave nanograting coupler},\ }\href
  {https://doi.org/https://doi.org/10.1038/ncomms3702} {\bibfield  {journal}
  {\bibinfo  {journal} {Nat. Comm.}\ }\textbf {\bibinfo {volume} {4}},\
  \bibinfo {pages} {2702} (\bibinfo {year} {2013})}\BibitemShut {NoStop}%
\bibitem [{\citenamefont {Salasyuk}\ \emph {et~al.}(2018)\citenamefont
  {Salasyuk}, \citenamefont {Rudkovskaya}, \citenamefont {Danilov},
  \citenamefont {Glavin}, \citenamefont {Kukhtaruk}, \citenamefont {Wang},
  \citenamefont {Rushforth}, \citenamefont {Nekludova}, \citenamefont
  {Sokolov}, \citenamefont {Elistratov}, \citenamefont {Yakovlev},
  \citenamefont {Bayer}, \citenamefont {Akimov},\ and\ \citenamefont
  {Scherbakov}}]{Salasyuk_2018}%
  \BibitemOpen
  \bibfield  {author} {\bibinfo {author} {\bibfnamefont {A.~S.}\ \bibnamefont
  {Salasyuk}}, \bibinfo {author} {\bibfnamefont {A.~V.}\ \bibnamefont
  {Rudkovskaya}}, \bibinfo {author} {\bibfnamefont {A.~P.}\ \bibnamefont
  {Danilov}}, \bibinfo {author} {\bibfnamefont {B.~A.}\ \bibnamefont {Glavin}},
  \bibinfo {author} {\bibfnamefont {S.~M.}\ \bibnamefont {Kukhtaruk}}, \bibinfo
  {author} {\bibfnamefont {M.}~\bibnamefont {Wang}}, \bibinfo {author}
  {\bibfnamefont {A.~W.}\ \bibnamefont {Rushforth}}, \bibinfo {author}
  {\bibfnamefont {P.~A.}\ \bibnamefont {Nekludova}}, \bibinfo {author}
  {\bibfnamefont {S.~V.}\ \bibnamefont {Sokolov}}, \bibinfo {author}
  {\bibfnamefont {A.~A.}\ \bibnamefont {Elistratov}}, \bibinfo {author}
  {\bibfnamefont {D.~R.}\ \bibnamefont {Yakovlev}}, \bibinfo {author}
  {\bibfnamefont {M.}~\bibnamefont {Bayer}}, \bibinfo {author} {\bibfnamefont
  {A.~V.}\ \bibnamefont {Akimov}},\ and\ \bibinfo {author} {\bibfnamefont
  {A.~V.}\ \bibnamefont {Scherbakov}},\ }\bibfield  {title} {\bibinfo {title}
  {Generation of a localized microwave magnetic field by coherent phonons in a
  ferromagnetic nanograting},\ }\href
  {https://doi.org/10.1103/physrevb.97.060404} {\bibfield  {journal} {\bibinfo
  {journal} {Phys. Rev. B}\ }\textbf {\bibinfo {volume} {97}},\ \bibinfo
  {pages} {060404(R)} (\bibinfo {year} {2018})}\BibitemShut {NoStop}%
\bibitem [{\citenamefont {Inoue}\ \emph {et~al.}(2011)\citenamefont {Inoue},
  \citenamefont {Baryshev}, \citenamefont {Takagi}, \citenamefont {Lim},
  \citenamefont {Hatafuku}, \citenamefont {Noda},\ and\ \citenamefont
  {Togo}}]{Inoue_2011}%
  \BibitemOpen
  \bibfield  {author} {\bibinfo {author} {\bibfnamefont {M.}~\bibnamefont
  {Inoue}}, \bibinfo {author} {\bibfnamefont {A.}~\bibnamefont {Baryshev}},
  \bibinfo {author} {\bibfnamefont {H.}~\bibnamefont {Takagi}}, \bibinfo
  {author} {\bibfnamefont {P.~B.}\ \bibnamefont {Lim}}, \bibinfo {author}
  {\bibfnamefont {K.}~\bibnamefont {Hatafuku}}, \bibinfo {author}
  {\bibfnamefont {J.}~\bibnamefont {Noda}},\ and\ \bibinfo {author}
  {\bibfnamefont {K.}~\bibnamefont {Togo}},\ }\bibfield  {title} {\bibinfo
  {title} {Investigating the use of magnonic crystals as extremely sensitive
  magnetic field sensors at room temperature},\ }\href
  {https://doi.org/10.1063/1.3567940} {\bibfield  {journal} {\bibinfo
  {journal} {Appl. Phys. Lett.}\ }\textbf {\bibinfo {volume} {98}},\ \bibinfo
  {pages} {132511} (\bibinfo {year} {2011})}\BibitemShut {NoStop}%
\bibitem [{\citenamefont {Bombeck}\ \emph {et~al.}(2012)\citenamefont
  {Bombeck}, \citenamefont {Salasyuk}, \citenamefont {Glavin}, \citenamefont
  {Scherbakov}, \citenamefont {Br\"uggemann}, \citenamefont {Yakovlev},
  \citenamefont {Sapega}, \citenamefont {Liu}, \citenamefont {Furdyna},
  \citenamefont {Akimov},\ and\ \citenamefont {Bayer}}]{Bombeck_2012}%
  \BibitemOpen
  \bibfield  {author} {\bibinfo {author} {\bibfnamefont {M.}~\bibnamefont
  {Bombeck}}, \bibinfo {author} {\bibfnamefont {A.~S.}\ \bibnamefont
  {Salasyuk}}, \bibinfo {author} {\bibfnamefont {B.~A.}\ \bibnamefont
  {Glavin}}, \bibinfo {author} {\bibfnamefont {A.~V.}\ \bibnamefont
  {Scherbakov}}, \bibinfo {author} {\bibfnamefont {C.}~\bibnamefont
  {Br\"uggemann}}, \bibinfo {author} {\bibfnamefont {D.~R.}\ \bibnamefont
  {Yakovlev}}, \bibinfo {author} {\bibfnamefont {V.~F.}\ \bibnamefont
  {Sapega}}, \bibinfo {author} {\bibfnamefont {X.}~\bibnamefont {Liu}},
  \bibinfo {author} {\bibfnamefont {J.~K.}\ \bibnamefont {Furdyna}}, \bibinfo
  {author} {\bibfnamefont {A.~V.}\ \bibnamefont {Akimov}},\ and\ \bibinfo
  {author} {\bibfnamefont {M.}~\bibnamefont {Bayer}},\ }\bibfield  {title}
  {\bibinfo {title} {Excitation of spin waves in ferromagnetic (ga,mn)as layers
  by picosecond strain pulses},\ }\href
  {https://doi.org/10.1103/PhysRevB.85.195324} {\bibfield  {journal} {\bibinfo
  {journal} {Phys. Rev. B}\ }\textbf {\bibinfo {volume} {85}},\ \bibinfo
  {pages} {195324} (\bibinfo {year} {2012})}\BibitemShut {NoStop}%
\bibitem [{\citenamefont {Verba}\ \emph {et~al.}(2018)\citenamefont {Verba},
  \citenamefont {Lisenkov}, \citenamefont {Krivorotov}, \citenamefont
  {Tiberkevich},\ and\ \citenamefont {Slavin}}]{Verba_2018}%
  \BibitemOpen
  \bibfield  {author} {\bibinfo {author} {\bibfnamefont {R.}~\bibnamefont
  {Verba}}, \bibinfo {author} {\bibfnamefont {I.}~\bibnamefont {Lisenkov}},
  \bibinfo {author} {\bibfnamefont {I.}~\bibnamefont {Krivorotov}}, \bibinfo
  {author} {\bibfnamefont {V.}~\bibnamefont {Tiberkevich}},\ and\ \bibinfo
  {author} {\bibfnamefont {A.}~\bibnamefont {Slavin}},\ }\bibfield  {title}
  {\bibinfo {title} {Nonreciprocal surface acoustic waves in multilayers with
  magnetoelastic and interfacial {Dzyaloshinskii-Moriya} interactions},\ }\href
  {https://doi.org/10.1103/physrevapplied.9.064014} {\bibfield  {journal}
  {\bibinfo  {journal} {Phys. Rev. Appl.}\ }\textbf {\bibinfo {volume} {9}},\
  \bibinfo {pages} {064014} (\bibinfo {year} {2018})}\BibitemShut {NoStop}%
\bibitem [{\citenamefont {Godejohann}\ \emph {et~al.}(2020)\citenamefont
  {Godejohann}, \citenamefont {Scherbakov}, \citenamefont {Kukhtaruk},
  \citenamefont {Poddubny}, \citenamefont {Yaremkevich}, \citenamefont {Wang},
  \citenamefont {Nadzeyka}, \citenamefont {Yakovlev}, \citenamefont
  {Rushforth}, \citenamefont {Akimov},\ and\ \citenamefont
  {Bayer}}]{Godejohann_2020}%
  \BibitemOpen
  \bibfield  {author} {\bibinfo {author} {\bibfnamefont {F.}~\bibnamefont
  {Godejohann}}, \bibinfo {author} {\bibfnamefont {A.~V.}\ \bibnamefont
  {Scherbakov}}, \bibinfo {author} {\bibfnamefont {S.~M.}\ \bibnamefont
  {Kukhtaruk}}, \bibinfo {author} {\bibfnamefont {A.~N.}\ \bibnamefont
  {Poddubny}}, \bibinfo {author} {\bibfnamefont {D.~D.}\ \bibnamefont
  {Yaremkevich}}, \bibinfo {author} {\bibfnamefont {M.}~\bibnamefont {Wang}},
  \bibinfo {author} {\bibfnamefont {A.}~\bibnamefont {Nadzeyka}}, \bibinfo
  {author} {\bibfnamefont {D.~R.}\ \bibnamefont {Yakovlev}}, \bibinfo {author}
  {\bibfnamefont {A.~W.}\ \bibnamefont {Rushforth}}, \bibinfo {author}
  {\bibfnamefont {A.~V.}\ \bibnamefont {Akimov}},\ and\ \bibinfo {author}
  {\bibfnamefont {M.}~\bibnamefont {Bayer}},\ }\bibfield  {title} {\bibinfo
  {title} {Magnon polaron formed by selectively coupled coherent magnon and
  phonon modes of a surface patterned ferromagnet},\ }\href
  {https://doi.org/10.1103/PhysRevB.102.144438} {\bibfield  {journal} {\bibinfo
   {journal} {Phys. Rev. B}\ }\textbf {\bibinfo {volume} {102}},\ \bibinfo
  {pages} {144438} (\bibinfo {year} {2020})}\BibitemShut {NoStop}%
\bibitem [{\citenamefont {Damon}\ and\ \citenamefont
  {Eshbach}(1961)}]{DE_1961}%
  \BibitemOpen
  \bibfield  {author} {\bibinfo {author} {\bibfnamefont {R.~W.}\ \bibnamefont
  {Damon}}\ and\ \bibinfo {author} {\bibfnamefont {J.~R.}\ \bibnamefont
  {Eshbach}},\ }\bibfield  {title} {\bibinfo {title} {Magnetostatic modes of a
  ferromagnet slab},\ }\href
  {https://doi.org/https://doi.org/10.1016/0022-3697(61)90041-5} {\bibfield
  {journal} {\bibinfo  {journal} {J. Phys. Chem. Solids}\ }\textbf {\bibinfo
  {volume} {19}},\ \bibinfo {pages} {308 } (\bibinfo {year}
  {1961})}\BibitemShut {NoStop}%
\bibitem [{\citenamefont {Schneider}\ \emph {et~al.}(2008)\citenamefont
  {Schneider}, \citenamefont {Serga}, \citenamefont {Neumann}, \citenamefont
  {Hillebrands},\ and\ \citenamefont {Kostylev}}]{Schneider_2008}%
  \BibitemOpen
  \bibfield  {author} {\bibinfo {author} {\bibfnamefont {T.}~\bibnamefont
  {Schneider}}, \bibinfo {author} {\bibfnamefont {A.~A.}\ \bibnamefont
  {Serga}}, \bibinfo {author} {\bibfnamefont {T.}~\bibnamefont {Neumann}},
  \bibinfo {author} {\bibfnamefont {B.}~\bibnamefont {Hillebrands}},\ and\
  \bibinfo {author} {\bibfnamefont {M.~P.}\ \bibnamefont {Kostylev}},\
  }\bibfield  {title} {\bibinfo {title} {Phase reciprocity of spin-wave
  excitation by a microstrip antenna},\ }\href
  {https://doi.org/10.1103/PhysRevB.77.214411} {\bibfield  {journal} {\bibinfo
  {journal} {Phys. Rev. B}\ }\textbf {\bibinfo {volume} {77}},\ \bibinfo
  {pages} {214411} (\bibinfo {year} {2008})}\BibitemShut {NoStop}%
\bibitem [{\citenamefont {Kostylev}(2013)}]{Kostylev_2013}%
  \BibitemOpen
  \bibfield  {author} {\bibinfo {author} {\bibfnamefont {M.}~\bibnamefont
  {Kostylev}},\ }\bibfield  {title} {\bibinfo {title} {Non-reciprocity of
  dipole-exchange spin waves in thin ferromagnetic films},\ }\href
  {https://doi.org/10.1063/1.4789962} {\bibfield  {journal} {\bibinfo
  {journal} {J. Appl. Phys.}\ }\textbf {\bibinfo {volume} {113}},\ \bibinfo
  {pages} {053907} (\bibinfo {year} {2013})}\BibitemShut {NoStop}%
\bibitem [{\citenamefont {Mruczkiewicz}\ \emph {et~al.}(2017)\citenamefont
  {Mruczkiewicz}, \citenamefont {Graczyk}, \citenamefont {Lupo}, \citenamefont
  {Adeyeye}, \citenamefont {Gubbiotti},\ and\ \citenamefont
  {Krawczyk}}]{Mruczkiewicz_2017}%
  \BibitemOpen
  \bibfield  {author} {\bibinfo {author} {\bibfnamefont {M.}~\bibnamefont
  {Mruczkiewicz}}, \bibinfo {author} {\bibfnamefont {P.}~\bibnamefont
  {Graczyk}}, \bibinfo {author} {\bibfnamefont {P.}~\bibnamefont {Lupo}},
  \bibinfo {author} {\bibfnamefont {A.}~\bibnamefont {Adeyeye}}, \bibinfo
  {author} {\bibfnamefont {G.}~\bibnamefont {Gubbiotti}},\ and\ \bibinfo
  {author} {\bibfnamefont {M.}~\bibnamefont {Krawczyk}},\ }\bibfield  {title}
  {\bibinfo {title} {Spin-wave nonreciprocity and magnonic band structure in a
  thin permalloy film induced by dynamical coupling with an array of ni
  stripes},\ }\href
  {https://doi.org/https://doi.org/10.1103/PhysRevB.96.104411} {\bibfield
  {journal} {\bibinfo  {journal} {Phys. Rev. B}\ }\textbf {\bibinfo {volume}
  {96}},\ \bibinfo {pages} {104411} (\bibinfo {year} {2017})}\BibitemShut
  {NoStop}%
\bibitem [{\citenamefont {Gubbiotti}\ \emph {et~al.}(2021)\citenamefont
  {Gubbiotti}, \citenamefont {Sadovnikov}, \citenamefont {Beginin},
  \citenamefont {Nikitov}, \citenamefont {Wan}, \citenamefont {Gupta},
  \citenamefont {Kundu}, \citenamefont {Talmelli}, \citenamefont {Carpenter},
  \citenamefont {Asselberghs}, \citenamefont {Radu}, \citenamefont {Adelmann},\
  and\ \citenamefont {Ciubotaru}}]{Gubbiotti_2021}%
  \BibitemOpen
  \bibfield  {author} {\bibinfo {author} {\bibfnamefont {G.}~\bibnamefont
  {Gubbiotti}}, \bibinfo {author} {\bibfnamefont {A.}~\bibnamefont
  {Sadovnikov}}, \bibinfo {author} {\bibfnamefont {E.}~\bibnamefont {Beginin}},
  \bibinfo {author} {\bibfnamefont {S.}~\bibnamefont {Nikitov}}, \bibinfo
  {author} {\bibfnamefont {D.}~\bibnamefont {Wan}}, \bibinfo {author}
  {\bibfnamefont {A.}~\bibnamefont {Gupta}}, \bibinfo {author} {\bibfnamefont
  {S.}~\bibnamefont {Kundu}}, \bibinfo {author} {\bibfnamefont
  {G.}~\bibnamefont {Talmelli}}, \bibinfo {author} {\bibfnamefont
  {R.}~\bibnamefont {Carpenter}}, \bibinfo {author} {\bibfnamefont
  {I.}~\bibnamefont {Asselberghs}}, \bibinfo {author} {\bibfnamefont {I.~P.}\
  \bibnamefont {Radu}}, \bibinfo {author} {\bibfnamefont {C.}~\bibnamefont
  {Adelmann}},\ and\ \bibinfo {author} {\bibfnamefont {F.}~\bibnamefont
  {Ciubotaru}},\ }\bibfield  {title} {\bibinfo {title} {Magnonic band structure
  in vertical meander-shaped
  ${\mathrm{co}}_{40}$${\mathrm{fe}}_{40}$${\mathrm{b}}_{20}$ thin films},\
  }\href {https://doi.org/10.1103/PhysRevApplied.15.014061} {\bibfield
  {journal} {\bibinfo  {journal} {Phys. Rev. Applied}\ }\textbf {\bibinfo
  {volume} {15}},\ \bibinfo {pages} {014061} (\bibinfo {year}
  {2021})}\BibitemShut {NoStop}%
\bibitem [{\citenamefont {Farle}(1998)}]{Farle_1998}%
  \BibitemOpen
  \bibfield  {author} {\bibinfo {author} {\bibfnamefont {M.}~\bibnamefont
  {Farle}},\ }\bibfield  {title} {\bibinfo {title} {Ferromagnetic resonance of
  ultrathin metallic layers},\ }\href
  {https://doi.org/10.1088/0034-4885/61/7/001} {\bibfield  {journal} {\bibinfo
  {journal} {Reports on Progress in Physics}\ }\textbf {\bibinfo {volume}
  {61}},\ \bibinfo {pages} {755} (\bibinfo {year} {1998})}\BibitemShut
  {NoStop}%
\bibitem [{\citenamefont {De~Wames}\ and\ \citenamefont
  {Wolfram}(1970)}]{Wames_1970}%
  \BibitemOpen
  \bibfield  {author} {\bibinfo {author} {\bibfnamefont {R.~E.}\ \bibnamefont
  {De~Wames}}\ and\ \bibinfo {author} {\bibfnamefont {T.}~\bibnamefont
  {Wolfram}},\ }\bibfield  {title} {\bibinfo {title} {Dipole-exchange spin
  waves in ferromagnetic films},\ }\href {https://doi.org/10.1063/1.1659049}
  {\bibfield  {journal} {\bibinfo  {journal} {Journal of Applied Physics}\
  }\textbf {\bibinfo {volume} {41}},\ \bibinfo {pages} {987} (\bibinfo {year}
  {1970})}\BibitemShut {NoStop}%
\bibitem [{\citenamefont {Hiebert}\ \emph {et~al.}(1997)\citenamefont
  {Hiebert}, \citenamefont {Stankiewicz},\ and\ \citenamefont
  {Freeman}}]{Hiebert_1997}%
  \BibitemOpen
  \bibfield  {author} {\bibinfo {author} {\bibfnamefont {W.~K.}\ \bibnamefont
  {Hiebert}}, \bibinfo {author} {\bibfnamefont {A.}~\bibnamefont
  {Stankiewicz}},\ and\ \bibinfo {author} {\bibfnamefont {M.~R.}\ \bibnamefont
  {Freeman}},\ }\bibfield  {title} {\bibinfo {title} {Direct observation of
  magnetic relaxation in a small permalloy disk by time-resolved scanning
  {Kerr} microscopy},\ }\href {https://doi.org/10.1103/physrevlett.79.1134}
  {\bibfield  {journal} {\bibinfo  {journal} {Phys. Rev. Lett.}\ }\textbf
  {\bibinfo {volume} {79}},\ \bibinfo {pages} {1134} (\bibinfo {year}
  {1997})}\BibitemShut {NoStop}%
\bibitem [{\citenamefont {van Kampen}\ \emph {et~al.}(2002)\citenamefont {van
  Kampen}, \citenamefont {Jozsa}, \citenamefont {Kohlhepp}, \citenamefont
  {LeClair}, \citenamefont {Lagae}, \citenamefont {de~Jonge},\ and\
  \citenamefont {Koopmans}}]{VanKampen_2002}%
  \BibitemOpen
  \bibfield  {author} {\bibinfo {author} {\bibfnamefont {M.}~\bibnamefont {van
  Kampen}}, \bibinfo {author} {\bibfnamefont {C.}~\bibnamefont {Jozsa}},
  \bibinfo {author} {\bibfnamefont {J.~T.}\ \bibnamefont {Kohlhepp}}, \bibinfo
  {author} {\bibfnamefont {P.}~\bibnamefont {LeClair}}, \bibinfo {author}
  {\bibfnamefont {L.}~\bibnamefont {Lagae}}, \bibinfo {author} {\bibfnamefont
  {W.~J.~M.}\ \bibnamefont {de~Jonge}},\ and\ \bibinfo {author} {\bibfnamefont
  {B.}~\bibnamefont {Koopmans}},\ }\bibfield  {title} {\bibinfo {title}
  {All-optical probe of coherent spin waves},\ }\href
  {https://doi.org/10.1103/PhysRevLett.88.227201} {\bibfield  {journal}
  {\bibinfo  {journal} {Phys. Rev. Lett.}\ }\textbf {\bibinfo {volume} {88}},\
  \bibinfo {pages} {227201} (\bibinfo {year} {2002})}\BibitemShut {NoStop}%
\bibitem [{\citenamefont {Kats}\ \emph {et~al.}(2016)\citenamefont {Kats},
  \citenamefont {Linnik}, \citenamefont {Salasyuk}, \citenamefont {Rushforth},
  \citenamefont {Wang}, \citenamefont {Wadley}, \citenamefont {Akimov},
  \citenamefont {Cavill}, \citenamefont {Holy}, \citenamefont {Kalashnikova},\
  and\ \citenamefont {Scherbakov}}]{Kats_2016}%
  \BibitemOpen
  \bibfield  {author} {\bibinfo {author} {\bibfnamefont {V.~N.}\ \bibnamefont
  {Kats}}, \bibinfo {author} {\bibfnamefont {T.~L.}\ \bibnamefont {Linnik}},
  \bibinfo {author} {\bibfnamefont {A.~S.}\ \bibnamefont {Salasyuk}}, \bibinfo
  {author} {\bibfnamefont {A.~W.}\ \bibnamefont {Rushforth}}, \bibinfo {author}
  {\bibfnamefont {M.}~\bibnamefont {Wang}}, \bibinfo {author} {\bibfnamefont
  {P.}~\bibnamefont {Wadley}}, \bibinfo {author} {\bibfnamefont {A.~V.}\
  \bibnamefont {Akimov}}, \bibinfo {author} {\bibfnamefont {S.~A.}\
  \bibnamefont {Cavill}}, \bibinfo {author} {\bibfnamefont {V.}~\bibnamefont
  {Holy}}, \bibinfo {author} {\bibfnamefont {A.~M.}\ \bibnamefont
  {Kalashnikova}},\ and\ \bibinfo {author} {\bibfnamefont {A.~V.}\ \bibnamefont
  {Scherbakov}},\ }\bibfield  {title} {\bibinfo {title} {Ultrafast changes of
  magnetic anisotropy driven by laser-generated coherent and noncoherent
  phonons in metallic films},\ }\href
  {https://doi.org/10.1103/PhysRevB.93.214422} {\bibfield  {journal} {\bibinfo
  {journal} {Phys. Rev. B}\ }\textbf {\bibinfo {volume} {93}},\ \bibinfo
  {pages} {214422} (\bibinfo {year} {2016})}\BibitemShut {NoStop}%
\bibitem [{\citenamefont {Scherbakov}\ \emph {et~al.}(2019)\citenamefont
  {Scherbakov}, \citenamefont {Danilov}, \citenamefont {Godejohann},
  \citenamefont {Linnik}, \citenamefont {Glavin}, \citenamefont {Shelukhin},
  \citenamefont {Pattnaik}, \citenamefont {Wang}, \citenamefont {Rushforth},
  \citenamefont {Yakovlev}, \citenamefont {Akimov},\ and\ \citenamefont
  {Bayer}}]{Scherbakov_2019}%
  \BibitemOpen
  \bibfield  {author} {\bibinfo {author} {\bibfnamefont {A.~V.}\ \bibnamefont
  {Scherbakov}}, \bibinfo {author} {\bibfnamefont {A.~P.}\ \bibnamefont
  {Danilov}}, \bibinfo {author} {\bibfnamefont {F.}~\bibnamefont {Godejohann}},
  \bibinfo {author} {\bibfnamefont {T.~L.}\ \bibnamefont {Linnik}}, \bibinfo
  {author} {\bibfnamefont {B.~A.}\ \bibnamefont {Glavin}}, \bibinfo {author}
  {\bibfnamefont {L.~A.}\ \bibnamefont {Shelukhin}}, \bibinfo {author}
  {\bibfnamefont {D.~P.}\ \bibnamefont {Pattnaik}}, \bibinfo {author}
  {\bibfnamefont {M.}~\bibnamefont {Wang}}, \bibinfo {author} {\bibfnamefont
  {A.~W.}\ \bibnamefont {Rushforth}}, \bibinfo {author} {\bibfnamefont {D.~R.}\
  \bibnamefont {Yakovlev}}, \bibinfo {author} {\bibfnamefont {A.~V.}\
  \bibnamefont {Akimov}},\ and\ \bibinfo {author} {\bibfnamefont
  {M.}~\bibnamefont {Bayer}},\ }\bibfield  {title} {\bibinfo {title} {Optical
  excitation of single- and multimode magnetization precession in
  $\mathrm{Fe}$-$\mathrm{Ga}$ nanolayers},\ }\href
  {https://doi.org/10.1103/PhysRevApplied.11.031003} {\bibfield  {journal}
  {\bibinfo  {journal} {Phys. Rev. Applied}\ }\textbf {\bibinfo {volume}
  {11}},\ \bibinfo {pages} {031003} (\bibinfo {year} {2019})}\BibitemShut
  {NoStop}%
\bibitem [{\citenamefont {Khokhlov}\ \emph {et~al.}(2019)\citenamefont
  {Khokhlov}, \citenamefont {Gerevenkov}, \citenamefont {Shelukhin},
  \citenamefont {Azovtsev}, \citenamefont {Pertsev}, \citenamefont {Wang},
  \citenamefont {Rushforth}, \citenamefont {Scherbakov},\ and\ \citenamefont
  {Kalashnikova}}]{Khokhlov_2019}%
  \BibitemOpen
  \bibfield  {author} {\bibinfo {author} {\bibfnamefont {N.~E.}\ \bibnamefont
  {Khokhlov}}, \bibinfo {author} {\bibfnamefont {P.~I.}\ \bibnamefont
  {Gerevenkov}}, \bibinfo {author} {\bibfnamefont {L.~A.}\ \bibnamefont
  {Shelukhin}}, \bibinfo {author} {\bibfnamefont {A.~V.}\ \bibnamefont
  {Azovtsev}}, \bibinfo {author} {\bibfnamefont {N.~A.}\ \bibnamefont
  {Pertsev}}, \bibinfo {author} {\bibfnamefont {M.}~\bibnamefont {Wang}},
  \bibinfo {author} {\bibfnamefont {A.~W.}\ \bibnamefont {Rushforth}}, \bibinfo
  {author} {\bibfnamefont {A.~V.}\ \bibnamefont {Scherbakov}},\ and\ \bibinfo
  {author} {\bibfnamefont {A.~M.}\ \bibnamefont {Kalashnikova}},\ }\bibfield
  {title} {\bibinfo {title} {Optical excitation of propagating magnetostatic
  waves in an epitaxial galfenol film by ultrafast magnetic anisotropy
  change},\ }\href {https://doi.org/10.1103/PhysRevApplied.12.044044}
  {\bibfield  {journal} {\bibinfo  {journal} {Phys. Rev. Applied}\ }\textbf
  {\bibinfo {volume} {12}},\ \bibinfo {pages} {044044} (\bibinfo {year}
  {2019})}\BibitemShut {NoStop}%
\bibitem [{Com()}]{Comsol}%
  \BibitemOpen
  \href {http://www.comsol.com} {}\bibinfo {note} {COMSOL
  Multiphysics$\textsuperscript{\textregistered}$~v.~5.4.~{\color{blue}
  www.comsol.com.} COMSOL AB, Stockholm, Sweden.}\BibitemShut {Stop}%
\bibitem [{\citenamefont {Lifshitz}\ and\ \citenamefont
  {Pitaevskii}(1980)}]{Landau_9}%
  \BibitemOpen
  \bibfield  {author} {\bibinfo {author} {\bibfnamefont {E.~M.}\ \bibnamefont
  {Lifshitz}}\ and\ \bibinfo {author} {\bibfnamefont {L.~P.}\ \bibnamefont
  {Pitaevskii}},\ }\href@noop {} {\emph {\bibinfo {title} {{Statistical
  Physics, Part 2: Theory of the Condensed State}}}},\ Vol.~\bibinfo {volume}
  {9}\ (\bibinfo  {publisher} {Butterworth-Heinemann},\ \bibinfo {year}
  {1980})\BibitemShut {NoStop}%
\bibitem [{\citenamefont {Bar'yakhtar}(1984)}]{LLB_1984}%
  \BibitemOpen
  \bibfield  {author} {\bibinfo {author} {\bibfnamefont {V.~G.}\ \bibnamefont
  {Bar'yakhtar}},\ }\bibfield  {title} {\bibinfo {title} {Phenomenological
  description of relaxation processes in magnetic materials},\ }\href
  {http://www.jetp.ac.ru/cgi-bin/dn/e_060_04_0863.pdf} {\bibfield  {journal}
  {\bibinfo  {journal} {Zh. Eksp. Teor. Fiz.}\ }\textbf {\bibinfo {volume}
  {87}},\ \bibinfo {pages} {1501} (\bibinfo {year} {1984})}\BibitemShut
  {NoStop}%
\bibitem [{\citenamefont {Chubykalo-Fesenko}\ \emph {et~al.}(2006)\citenamefont
  {Chubykalo-Fesenko}, \citenamefont {Nowak}, \citenamefont {Chantrell},\ and\
  \citenamefont {Garanin}}]{Fesenko_2006}%
  \BibitemOpen
  \bibfield  {author} {\bibinfo {author} {\bibfnamefont {O.}~\bibnamefont
  {Chubykalo-Fesenko}}, \bibinfo {author} {\bibfnamefont {U.}~\bibnamefont
  {Nowak}}, \bibinfo {author} {\bibfnamefont {R.~W.}\ \bibnamefont
  {Chantrell}},\ and\ \bibinfo {author} {\bibfnamefont {D.}~\bibnamefont
  {Garanin}},\ }\bibfield  {title} {\bibinfo {title} {Dynamic approach for
  micromagnetics close to the curie temperature},\ }\href
  {https://doi.org/10.1103/PhysRevB.74.094436} {\bibfield  {journal} {\bibinfo
  {journal} {Phys. Rev. B}\ }\textbf {\bibinfo {volume} {74}},\ \bibinfo
  {pages} {094436} (\bibinfo {year} {2006})}\BibitemShut {NoStop}%
\bibitem [{SM()}]{SM}%
  \BibitemOpen
  \href@noop {} {}\bibinfo {note} {See Supplemental Material below for details
  related to the static demagnetizing field of the nanogratings, magnons in
  nanogratings for DE and BV geometries, spatial Fourier transform of magnon
  modes, and the verification of micromagnetic simulations using COMSOL
  Multiphysics.}\BibitemShut {Stop}%
\bibitem [{Two()}]{TwoNumbers}%
  \BibitemOpen
  \href@noop {} {}\bibinfo {note} {Alternatively, the Fourier transform can be
  performed separately in the wire and groove regions. In this case there are
  two "magnon quantum numbers" to characterize the
  magnons~\cite{Langer_2019}.}\BibitemShut {Stop}%
\bibitem [{\citenamefont {Yang}\ \emph {et~al.}(2019)\citenamefont {Yang},
  \citenamefont {Jaris}, \citenamefont {Berk},\ and\ \citenamefont
  {Schmidt}}]{Yang_2019}%
  \BibitemOpen
  \bibfield  {author} {\bibinfo {author} {\bibfnamefont {W.~G.}\ \bibnamefont
  {Yang}}, \bibinfo {author} {\bibfnamefont {M.}~\bibnamefont {Jaris}},
  \bibinfo {author} {\bibfnamefont {C.}~\bibnamefont {Berk}},\ and\ \bibinfo
  {author} {\bibfnamefont {H.}~\bibnamefont {Schmidt}},\ }\bibfield  {title}
  {\bibinfo {title} {Preferential excitation of a single nanomagnet using
  magnetoelastic coupling},\ }\href
  {https://doi.org/10.1103/PhysRevB.99.104434} {\bibfield  {journal} {\bibinfo
  {journal} {Phys. Rev. B}\ }\textbf {\bibinfo {volume} {99}},\ \bibinfo
  {pages} {104434} (\bibinfo {year} {2019})}\BibitemShut {NoStop}%
\bibitem [{\citenamefont {Babu}\ \emph {et~al.}(2021)\citenamefont {Babu},
  \citenamefont {Trzaskowska}, \citenamefont {Graczyk}, \citenamefont
  {Centa\l{}a}, \citenamefont {Mieszczak}, \citenamefont {G\l{}owi\'{n}ski},
  \citenamefont {Zdunek}, \citenamefont {Mielcarek},\ and\ \citenamefont
  {K\l{}os}}]{Babu_2021}%
  \BibitemOpen
  \bibfield  {author} {\bibinfo {author} {\bibfnamefont {N.~K.~P.}\
  \bibnamefont {Babu}}, \bibinfo {author} {\bibfnamefont {A.}~\bibnamefont
  {Trzaskowska}}, \bibinfo {author} {\bibfnamefont {P.}~\bibnamefont
  {Graczyk}}, \bibinfo {author} {\bibfnamefont {G.}~\bibnamefont {Centa\l{}a}},
  \bibinfo {author} {\bibfnamefont {S.}~\bibnamefont {Mieszczak}}, \bibinfo
  {author} {\bibfnamefont {H.}~\bibnamefont {G\l{}owi\'{n}ski}}, \bibinfo
  {author} {\bibfnamefont {M.}~\bibnamefont {Zdunek}}, \bibinfo {author}
  {\bibfnamefont {S.}~\bibnamefont {Mielcarek}},\ and\ \bibinfo {author}
  {\bibfnamefont {J.~W.}\ \bibnamefont {K\l{}os}},\ }\bibfield  {title}
  {\bibinfo {title} {The interaction between surface acoustic waves and spin
  waves: The role of anisotropy and spatial profiles of the modes},\ }\href
  {https://doi.org/10.1021/acs.nanolett.0c03692} {\bibfield  {journal}
  {\bibinfo  {journal} {Nano Letters}\ }\textbf {\bibinfo {volume} {21}},\
  \bibinfo {pages} {946} (\bibinfo {year} {2021})}\BibitemShut {NoStop}%
\bibitem [{\citenamefont {Kalinikos}\ and\ \citenamefont
  {Slavin}(1986)}]{Kalinikos_1986}%
  \BibitemOpen
  \bibfield  {author} {\bibinfo {author} {\bibfnamefont {B.~A.}\ \bibnamefont
  {Kalinikos}}\ and\ \bibinfo {author} {\bibfnamefont {A.~N.}\ \bibnamefont
  {Slavin}},\ }\bibfield  {title} {\bibinfo {title} {Theory of dipole-exchange
  spin wave spectrum for ferromagnetic films with mixed exchange boundary
  conditions},\ }\href {https://doi.org/10.1088/0022-3719/19/35/014} {\bibfield
   {journal} {\bibinfo  {journal} {Journal of Physics C: Solid State Physics}\
  }\textbf {\bibinfo {volume} {19}},\ \bibinfo {pages} {7013} (\bibinfo {year}
  {1986})}\BibitemShut {NoStop}%
\bibitem [{\citenamefont {Gallardo}\ \emph {et~al.}(2014)\citenamefont
  {Gallardo}, \citenamefont {Banholzer}, \citenamefont {Wagner}, \citenamefont
  {K\"{o}rner}, \citenamefont {Lenz}, \citenamefont {Farle}, \citenamefont
  {Lindner}, \citenamefont {Fassbender},\ and\ \citenamefont
  {Landeros}}]{Gallardo_2014}%
  \BibitemOpen
  \bibfield  {author} {\bibinfo {author} {\bibfnamefont {R.~A.}\ \bibnamefont
  {Gallardo}}, \bibinfo {author} {\bibfnamefont {A.}~\bibnamefont {Banholzer}},
  \bibinfo {author} {\bibfnamefont {K.}~\bibnamefont {Wagner}}, \bibinfo
  {author} {\bibfnamefont {M.}~\bibnamefont {K\"{o}rner}}, \bibinfo {author}
  {\bibfnamefont {K.}~\bibnamefont {Lenz}}, \bibinfo {author} {\bibfnamefont
  {M.}~\bibnamefont {Farle}}, \bibinfo {author} {\bibfnamefont
  {J.}~\bibnamefont {Lindner}}, \bibinfo {author} {\bibfnamefont
  {J.}~\bibnamefont {Fassbender}},\ and\ \bibinfo {author} {\bibfnamefont
  {P.}~\bibnamefont {Landeros}},\ }\bibfield  {title} {\bibinfo {title}
  {Splitting of spin-wave modes in thin films with arrays of periodic
  perturbations: theory and experiment},\ }\href
  {https://doi.org/10.1088/1367-2630/16/2/023015} {\bibfield  {journal}
  {\bibinfo  {journal} {New Journal of Physics}\ }\textbf {\bibinfo {volume}
  {16}},\ \bibinfo {pages} {023015} (\bibinfo {year} {2014})}\BibitemShut
  {NoStop}%
\bibitem [{\citenamefont {$\textstyle{\rm \mu MAG}$}()}]{SMMP}%
  \BibitemOpen
  \bibfield  {author} {\bibinfo {author} {\bibnamefont {$\textstyle{\rm \mu
  MAG}$}},\ }\href {https://www.ctcms.nist.gov/~rdm/mumag.org.html} {\bibinfo
  {journal} {{\color{black} Micromagnetic Modeling Activity Group,}
  www.ctcms.nist.gov/$\sim$rdm/mumag.org.html$\!\!$}\ }\BibitemShut {NoStop}%
\bibitem [{\citenamefont {Wang}\ \emph {et~al.}(2017)\citenamefont {Wang},
  \citenamefont {Li}, \citenamefont {Liang}, \citenamefont {Barra},
  \citenamefont {Domann}, \citenamefont {Lynch}, \citenamefont {Sepulveda},\
  and\ \citenamefont {Carman}}]{CoFeB_le}%
  \BibitemOpen
\bibfield  {journal} {  }\bibfield  {author} {\bibinfo {author} {\bibfnamefont
  {Q.}~\bibnamefont {Wang}}, \bibinfo {author} {\bibfnamefont {X.}~\bibnamefont
  {Li}}, \bibinfo {author} {\bibfnamefont {C.-Y.}\ \bibnamefont {Liang}},
  \bibinfo {author} {\bibfnamefont {A.}~\bibnamefont {Barra}}, \bibinfo
  {author} {\bibfnamefont {J.}~\bibnamefont {Domann}}, \bibinfo {author}
  {\bibfnamefont {C.}~\bibnamefont {Lynch}}, \bibinfo {author} {\bibfnamefont
  {A.}~\bibnamefont {Sepulveda}},\ and\ \bibinfo {author} {\bibfnamefont
  {G.}~\bibnamefont {Carman}},\ }\bibfield  {title} {\bibinfo {title}
  {Strain-mediated $180^{\circ}$ switching in $\rm cofeb$ and terfenol-d
  nanodots with perpendicular magnetic anisotropy},\ }\href
  {https://doi.org/10.1063/1.4978270} {\bibfield  {journal} {\bibinfo
  {journal} {Applied Physics Letters}\ }\textbf {\bibinfo {volume} {110}},\
  \bibinfo {pages} {102903} (\bibinfo {year} {2017})}\BibitemShut {NoStop}%
\bibitem [{\citenamefont {Manna}\ \emph {et~al.}(2017)\citenamefont {Manna},
  \citenamefont {Kim}, \citenamefont {Lubarda}, \citenamefont {Wingert},
  \citenamefont {Harder}, \citenamefont {Spada}, \citenamefont {Lomakin},
  \citenamefont {Shpyrko},\ and\ \citenamefont {Fullerton}}]{Ni_le}%
  \BibitemOpen
  \bibfield  {author} {\bibinfo {author} {\bibfnamefont {S.}~\bibnamefont
  {Manna}}, \bibinfo {author} {\bibfnamefont {J.~W.}\ \bibnamefont {Kim}},
  \bibinfo {author} {\bibfnamefont {M.~V.}\ \bibnamefont {Lubarda}}, \bibinfo
  {author} {\bibfnamefont {J.}~\bibnamefont {Wingert}}, \bibinfo {author}
  {\bibfnamefont {R.}~\bibnamefont {Harder}}, \bibinfo {author} {\bibfnamefont
  {F.}~\bibnamefont {Spada}}, \bibinfo {author} {\bibfnamefont
  {V.}~\bibnamefont {Lomakin}}, \bibinfo {author} {\bibfnamefont
  {O.}~\bibnamefont {Shpyrko}},\ and\ \bibinfo {author} {\bibfnamefont {E.~E.}\
  \bibnamefont {Fullerton}},\ }\bibfield  {title} {\bibinfo {title}
  {Characterization of strain and its effects on ferromagnetic nickel
  nanocubes},\ }\href {https://doi.org/10.1063/1.5004577} {\bibfield  {journal}
  {\bibinfo  {journal} {AIP Advances}\ }\textbf {\bibinfo {volume} {7}},\
  \bibinfo {pages} {125025} (\bibinfo {year} {2017})}\BibitemShut {NoStop}%
\bibitem [{\citenamefont {Bautin}\ \emph {et~al.}(2017)\citenamefont {Bautin},
  \citenamefont {Seferyan}, \citenamefont {Nesmeyanov},\ and\ \citenamefont
  {Usov}}]{Co_le}%
  \BibitemOpen
  \bibfield  {author} {\bibinfo {author} {\bibfnamefont {V.~A.}\ \bibnamefont
  {Bautin}}, \bibinfo {author} {\bibfnamefont {A.~G.}\ \bibnamefont
  {Seferyan}}, \bibinfo {author} {\bibfnamefont {M.~S.}\ \bibnamefont
  {Nesmeyanov}},\ and\ \bibinfo {author} {\bibfnamefont {N.~A.}\ \bibnamefont
  {Usov}},\ }\bibfield  {title} {\bibinfo {title} {Magnetic properties of
  polycrystalline cobalt nanoparticles},\ }\href
  {https://doi.org/10.1063/1.4979889} {\bibfield  {journal} {\bibinfo
  {journal} {AIP Advances}\ }\textbf {\bibinfo {volume} {7}},\ \bibinfo {pages}
  {045103} (\bibinfo {year} {2017})}\BibitemShut {NoStop}%
\bibitem [{\citenamefont {Piao}\ \emph {et~al.}(2011)\citenamefont {Piao},
  \citenamefont {Choi}, \citenamefont {Shim}, \citenamefont {Kim},\ and\
  \citenamefont {You}}]{Piao_2011}%
  \BibitemOpen
  \bibfield  {author} {\bibinfo {author} {\bibfnamefont {H.-G.}\ \bibnamefont
  {Piao}}, \bibinfo {author} {\bibfnamefont {H.-C.}\ \bibnamefont {Choi}},
  \bibinfo {author} {\bibfnamefont {J.-H.}\ \bibnamefont {Shim}}, \bibinfo
  {author} {\bibfnamefont {D.-H.}\ \bibnamefont {Kim}},\ and\ \bibinfo {author}
  {\bibfnamefont {C.-Y.}\ \bibnamefont {You}},\ }\bibfield  {title} {\bibinfo
  {title} {Ratchet effect of the domain wall by asymmetric magnetostatic
  potentials},\ }\href {https://doi.org/10.1063/1.3658805} {\bibfield
  {journal} {\bibinfo  {journal} {Applied Physics Letters}\ }\textbf {\bibinfo
  {volume} {99}},\ \bibinfo {pages} {192512} (\bibinfo {year}
  {2011})}\BibitemShut {NoStop}%
\bibitem [{\citenamefont {Gopman}\ \emph {et~al.}(2017)\citenamefont {Gopman},
  \citenamefont {Sampath}, \citenamefont {Ahmad}, \citenamefont
  {Bandyopadhyay},\ and\ \citenamefont {Atulasimha}}]{Gopman_2017}%
  \BibitemOpen
  \bibfield  {author} {\bibinfo {author} {\bibfnamefont {D.~B.}\ \bibnamefont
  {Gopman}}, \bibinfo {author} {\bibfnamefont {V.}~\bibnamefont {Sampath}},
  \bibinfo {author} {\bibfnamefont {H.}~\bibnamefont {Ahmad}}, \bibinfo
  {author} {\bibfnamefont {S.}~\bibnamefont {Bandyopadhyay}},\ and\ \bibinfo
  {author} {\bibfnamefont {J.}~\bibnamefont {Atulasimha}},\ }\bibfield  {title}
  {\bibinfo {title} {Static and dynamic magnetic properties of sputtered
  {Fe-Ga} thin films},\ }\href {https://doi.org/10.1109/tmag.2017.2700404}
  {\bibfield  {journal} {\bibinfo  {journal} {{IEEE} Trans. Mag.}\ }\textbf
  {\bibinfo {volume} {53}},\ \bibinfo {pages} {1} (\bibinfo {year}
  {2017})}\BibitemShut {NoStop}%
\bibitem [{\citenamefont {Klingler}\ \emph {et~al.}(2014)\citenamefont
  {Klingler}, \citenamefont {Chumak}, \citenamefont {Mewes}, \citenamefont
  {Khodadadi}, \citenamefont {Mewes}, \citenamefont {Dubs}, \citenamefont
  {Surzhenko}, \citenamefont {Hillebrands},\ and\ \citenamefont
  {Conca}}]{Klingler_2014}%
  \BibitemOpen
  \bibfield  {author} {\bibinfo {author} {\bibfnamefont {S.}~\bibnamefont
  {Klingler}}, \bibinfo {author} {\bibfnamefont {A.~V.}\ \bibnamefont
  {Chumak}}, \bibinfo {author} {\bibfnamefont {T.}~\bibnamefont {Mewes}},
  \bibinfo {author} {\bibfnamefont {B.}~\bibnamefont {Khodadadi}}, \bibinfo
  {author} {\bibfnamefont {C.}~\bibnamefont {Mewes}}, \bibinfo {author}
  {\bibfnamefont {C.}~\bibnamefont {Dubs}}, \bibinfo {author} {\bibfnamefont
  {O.}~\bibnamefont {Surzhenko}}, \bibinfo {author} {\bibfnamefont
  {B.}~\bibnamefont {Hillebrands}},\ and\ \bibinfo {author} {\bibfnamefont
  {A.}~\bibnamefont {Conca}},\ }\bibfield  {title} {\bibinfo {title}
  {Measurements of the exchange stiffness of {YIG} films using broadband
  ferromagnetic resonance techniques},\ }\href
  {https://doi.org/10.1088/0022-3727/48/1/015001} {\bibfield  {journal}
  {\bibinfo  {journal} {Journal of Physics D: Applied Physics}\ }\textbf
  {\bibinfo {volume} {48}},\ \bibinfo {pages} {015001} (\bibinfo {year}
  {2014})}\BibitemShut {NoStop}%
\bibitem [{\citenamefont {Awschalom}\ \emph {et~al.}(2021)\citenamefont
  {Awschalom}, \citenamefont {Du}, \citenamefont {He}, \citenamefont
  {Heremans}, \citenamefont {Hoffmann}, \citenamefont {Hou}, \citenamefont
  {Kurebayashi}, \citenamefont {Li}, \citenamefont {Liu}, \citenamefont
  {Novosad}, \citenamefont {Sklenar}, \citenamefont {Sullivan}, \citenamefont
  {Sun}, \citenamefont {Tang}, \citenamefont {Tyberkevych}, \citenamefont
  {Trevillian}, \citenamefont {Tsen}, \citenamefont {Weiss}, \citenamefont
  {Zhang}, \citenamefont {Zhang}, \citenamefont {Zhao},\ and\ \citenamefont
  {Zollitsch}}]{Awschalom_2021}%
  \BibitemOpen
  \bibfield  {author} {\bibinfo {author} {\bibfnamefont {D.~D.}\ \bibnamefont
  {Awschalom}}, \bibinfo {author} {\bibfnamefont {C.~R.}\ \bibnamefont {Du}},
  \bibinfo {author} {\bibfnamefont {R.}~\bibnamefont {He}}, \bibinfo {author}
  {\bibfnamefont {F.~J.}\ \bibnamefont {Heremans}}, \bibinfo {author}
  {\bibfnamefont {A.}~\bibnamefont {Hoffmann}}, \bibinfo {author}
  {\bibfnamefont {J.}~\bibnamefont {Hou}}, \bibinfo {author} {\bibfnamefont
  {H.}~\bibnamefont {Kurebayashi}}, \bibinfo {author} {\bibfnamefont
  {Y.}~\bibnamefont {Li}}, \bibinfo {author} {\bibfnamefont {L.}~\bibnamefont
  {Liu}}, \bibinfo {author} {\bibfnamefont {V.}~\bibnamefont {Novosad}},
  \bibinfo {author} {\bibfnamefont {J.}~\bibnamefont {Sklenar}}, \bibinfo
  {author} {\bibfnamefont {S.~E.}\ \bibnamefont {Sullivan}}, \bibinfo {author}
  {\bibfnamefont {D.}~\bibnamefont {Sun}}, \bibinfo {author} {\bibfnamefont
  {H.}~\bibnamefont {Tang}}, \bibinfo {author} {\bibfnamefont {V.}~\bibnamefont
  {Tyberkevych}}, \bibinfo {author} {\bibfnamefont {C.}~\bibnamefont
  {Trevillian}}, \bibinfo {author} {\bibfnamefont {A.~W.}\ \bibnamefont
  {Tsen}}, \bibinfo {author} {\bibfnamefont {L.~R.}\ \bibnamefont {Weiss}},
  \bibinfo {author} {\bibfnamefont {W.}~\bibnamefont {Zhang}}, \bibinfo
  {author} {\bibfnamefont {X.}~\bibnamefont {Zhang}}, \bibinfo {author}
  {\bibfnamefont {L.}~\bibnamefont {Zhao}},\ and\ \bibinfo {author}
  {\bibfnamefont {C.~W.}\ \bibnamefont {Zollitsch}},\ }\bibfield  {title}
  {\bibinfo {title} {Quantum engineering with hybrid magnonic systems and
  materials},\ }\href {https://doi.org/10.1109/TQE.2021.3057799} {\bibfield
  {journal} {\bibinfo  {journal} {IEEE Transactions on Quantum Engineering}\
  }\textbf {\bibinfo {volume} {2}},\ \bibinfo {pages} {1} (\bibinfo {year}
  {2021})}\BibitemShut {NoStop}%
\bibitem [{\citenamefont {Li}\ \emph {et~al.}(2020)\citenamefont {Li},
  \citenamefont {Zhang}, \citenamefont {Tyberkevych}, \citenamefont {Kwok},
  \citenamefont {Hoffmann},\ and\ \citenamefont {Novosad}}]{Li-2020}%
  \BibitemOpen
  \bibfield  {author} {\bibinfo {author} {\bibfnamefont {Y.}~\bibnamefont
  {Li}}, \bibinfo {author} {\bibfnamefont {W.}~\bibnamefont {Zhang}}, \bibinfo
  {author} {\bibfnamefont {V.}~\bibnamefont {Tyberkevych}}, \bibinfo {author}
  {\bibfnamefont {W.-K.}\ \bibnamefont {Kwok}}, \bibinfo {author}
  {\bibfnamefont {A.}~\bibnamefont {Hoffmann}},\ and\ \bibinfo {author}
  {\bibfnamefont {V.}~\bibnamefont {Novosad}},\ }\bibfield  {title} {\bibinfo
  {title} {Hybrid magnonics: Physics, circuits, and applications for coherent
  information processing},\ }\href {https://doi.org/10.1063/5.0020277}
  {\bibfield  {journal} {\bibinfo  {journal} {J. Appl. Phys.}\ }\textbf
  {\bibinfo {volume} {128}},\ \bibinfo {pages} {130902} (\bibinfo {year}
  {2020})}\BibitemShut {NoStop}%
\bibitem [{\citenamefont {Satoh}\ \emph {et~al.}(2012)\citenamefont {Satoh},
  \citenamefont {Terui}, \citenamefont {Moriya}, \citenamefont {Ivanov},
  \citenamefont {Ando}, \citenamefont {Saitoh}, \citenamefont {Shimura},\ and\
  \citenamefont {Kuroda}}]{Ivanov_2012}%
  \BibitemOpen
  \bibfield  {author} {\bibinfo {author} {\bibfnamefont {T.}~\bibnamefont
  {Satoh}}, \bibinfo {author} {\bibfnamefont {Y.}~\bibnamefont {Terui}},
  \bibinfo {author} {\bibfnamefont {R.}~\bibnamefont {Moriya}}, \bibinfo
  {author} {\bibfnamefont {B.~A.}\ \bibnamefont {Ivanov}}, \bibinfo {author}
  {\bibfnamefont {K.}~\bibnamefont {Ando}}, \bibinfo {author} {\bibfnamefont
  {E.}~\bibnamefont {Saitoh}}, \bibinfo {author} {\bibfnamefont
  {T.}~\bibnamefont {Shimura}},\ and\ \bibinfo {author} {\bibfnamefont
  {K.}~\bibnamefont {Kuroda}},\ }\bibfield  {title} {\bibinfo {title}
  {Directional control of spin-wave emission by spatially shaped light},\
  }\href {https://doi.org/https://doi.org/10.1038/nphoton.2012.218} {\bibfield
  {journal} {\bibinfo  {journal} {Nat. Photonics}\ }\textbf {\bibinfo {volume}
  {6}},\ \bibinfo {pages} {662} (\bibinfo {year} {2012})}\BibitemShut {NoStop}%
\bibitem [{\citenamefont {J\"ackl}\ \emph {et~al.}(2017)\citenamefont
  {J\"ackl}, \citenamefont {Belotelov}, \citenamefont {Akimov}, \citenamefont
  {Savochkin}, \citenamefont {Yakovlev}, \citenamefont {Zvezdin},\ and\
  \citenamefont {Bayer}}]{Jackl_2017}%
  \BibitemOpen
  \bibfield  {author} {\bibinfo {author} {\bibfnamefont {M.}~\bibnamefont
  {J\"ackl}}, \bibinfo {author} {\bibfnamefont {V.~I.}\ \bibnamefont
  {Belotelov}}, \bibinfo {author} {\bibfnamefont {I.~A.}\ \bibnamefont
  {Akimov}}, \bibinfo {author} {\bibfnamefont {I.~V.}\ \bibnamefont
  {Savochkin}}, \bibinfo {author} {\bibfnamefont {D.~R.}\ \bibnamefont
  {Yakovlev}}, \bibinfo {author} {\bibfnamefont {A.~K.}\ \bibnamefont
  {Zvezdin}},\ and\ \bibinfo {author} {\bibfnamefont {M.}~\bibnamefont
  {Bayer}},\ }\bibfield  {title} {\bibinfo {title} {Magnon accumulation by
  clocked laser excitation as source of long-range spin waves in transparent
  magnetic films},\ }\href {https://doi.org/10.1103/PhysRevX.7.021009}
  {\bibfield  {journal} {\bibinfo  {journal} {Phys. Rev. X}\ }\textbf {\bibinfo
  {volume} {7}},\ \bibinfo {pages} {021009} (\bibinfo {year}
  {2017})}\BibitemShut {NoStop}%
\bibitem [{\citenamefont {Schoen}\ \emph {et~al.}(2016)\citenamefont {Schoen},
  \citenamefont {Thonig}, \citenamefont {Schneider}, \citenamefont {Silva},
  \citenamefont {Nembach}, \citenamefont {Eriksson}, \citenamefont {Karis},\
  and\ \citenamefont {Shaw}}]{Schoen_2016}%
  \BibitemOpen
  \bibfield  {author} {\bibinfo {author} {\bibfnamefont {M.~A.~W.}\
  \bibnamefont {Schoen}}, \bibinfo {author} {\bibfnamefont {D.}~\bibnamefont
  {Thonig}}, \bibinfo {author} {\bibfnamefont {M.~L.}\ \bibnamefont
  {Schneider}}, \bibinfo {author} {\bibfnamefont {T.~J.}\ \bibnamefont
  {Silva}}, \bibinfo {author} {\bibfnamefont {H.~T.}\ \bibnamefont {Nembach}},
  \bibinfo {author} {\bibfnamefont {O.}~\bibnamefont {Eriksson}}, \bibinfo
  {author} {\bibfnamefont {O.}~\bibnamefont {Karis}},\ and\ \bibinfo {author}
  {\bibfnamefont {J.~M.}\ \bibnamefont {Shaw}},\ }\bibfield  {title} {\bibinfo
  {title} {Ultra-low magnetic damping of a metallic ferromagnet},\ }\href
  {https://doi.org/https://doi.org/10.1038/nphys3770} {\bibfield  {journal}
  {\bibinfo  {journal} {Nat. Phys.}\ }\textbf {\bibinfo {volume} {21}},\
  \bibinfo {pages} {839} (\bibinfo {year} {2016})}\BibitemShut {NoStop}%
\bibitem [{\citenamefont {Yu}\ \emph {et~al.}(2012)\citenamefont {Yu},
  \citenamefont {Huber}, \citenamefont {Schwarze}, \citenamefont {Brandl},
  \citenamefont {Rapp}, \citenamefont {Berberich}, \citenamefont {Duerr},\ and\
  \citenamefont {Grundler}}]{Yu_2012}%
  \BibitemOpen
  \bibfield  {author} {\bibinfo {author} {\bibfnamefont {H.}~\bibnamefont
  {Yu}}, \bibinfo {author} {\bibfnamefont {R.}~\bibnamefont {Huber}}, \bibinfo
  {author} {\bibfnamefont {T.}~\bibnamefont {Schwarze}}, \bibinfo {author}
  {\bibfnamefont {F.}~\bibnamefont {Brandl}}, \bibinfo {author} {\bibfnamefont
  {T.}~\bibnamefont {Rapp}}, \bibinfo {author} {\bibfnamefont {P.}~\bibnamefont
  {Berberich}}, \bibinfo {author} {\bibfnamefont {G.}~\bibnamefont {Duerr}},\
  and\ \bibinfo {author} {\bibfnamefont {D.}~\bibnamefont {Grundler}},\
  }\bibfield  {title} {\bibinfo {title} {High propagating velocity of spin
  waves and temperature dependent damping in a cofeb thin film},\ }\href
  {https://doi.org/10.1063/1.4731273} {\bibfield  {journal} {\bibinfo
  {journal} {Appl. Phys. Lett.}\ }\textbf {\bibinfo {volume} {100}},\ \bibinfo
  {pages} {262412} (\bibinfo {year} {2012})}\BibitemShut {NoStop}%
\end{thebibliography}%


\begin{thebibliography}{11}%
\makeatletter
\providecommand \@ifxundefined [1]{%
 \@ifx{#1\undefined}
}%
\providecommand \@ifnum [1]{%
 \ifnum #1\expandafter \@firstoftwo
 \else \expandafter \@secondoftwo
 \fi
}%
\providecommand \@ifx [1]{%
 \ifx #1\expandafter \@firstoftwo
 \else \expandafter \@secondoftwo
 \fi
}%
\providecommand \natexlab [1]{#1}%
\providecommand \enquote  [1]{``#1''}%
\providecommand \bibnamefont  [1]{#1}%
\providecommand \bibfnamefont [1]{#1}%
\providecommand \citenamefont [1]{#1}%
\providecommand \href@noop [0]{\@secondoftwo}%
\providecommand \href [0]{\begingroup \@sanitize@url \@href}%
\providecommand \@href[1]{\@@startlink{#1}\@@href}%
\providecommand \@@href[1]{\endgroup#1\@@endlink}%
\providecommand \@sanitize@url [0]{\catcode `\\12\catcode `\$12\catcode
  `\&12\catcode `\#12\catcode `\^12\catcode `\_12\catcode `\%12\relax}%
\providecommand \@@startlink[1]{}%
\providecommand \@@endlink[0]{}%
\providecommand \url  [0]{\begingroup\@sanitize@url \@url }%
\providecommand \@url [1]{\endgroup\@href {#1}{\urlprefix }}%
\providecommand \urlprefix  [0]{URL }%
\providecommand \Eprint [0]{\href }%
\providecommand \doibase [0]{https://doi.org/}%
\providecommand \selectlanguage [0]{\@gobble}%
\providecommand \bibinfo  [0]{\@secondoftwo}%
\providecommand \bibfield  [0]{\@secondoftwo}%
\providecommand \translation [1]{[#1]}%
\providecommand \BibitemOpen [0]{}%
\providecommand \bibitemStop [0]{}%
\providecommand \bibitemNoStop [0]{.\EOS\space}%
\providecommand \EOS [0]{\spacefactor3000\relax}%
\providecommand \BibitemShut  [1]{\csname bibitem#1\endcsname}%
\let\auto@bib@innerbib\@empty
\bibitem [{\citenamefont {Langer}\ \emph {et~al.}(2019)\citenamefont {Langer},
  \citenamefont {Gallardo}, \citenamefont {Schneider}, \citenamefont {Stienen},
  \citenamefont {Rold{\'{a}}n-Molina}, \citenamefont {Yuan}, \citenamefont
  {Lenz}, \citenamefont {Lindner}, \citenamefont {Landeros},\ and\
  \citenamefont {Fassbender}}]{Langer_2019}%
  \BibitemOpen
  \bibfield  {author} {\bibinfo {author} {\bibfnamefont {M.}~\bibnamefont
  {Langer}}, \bibinfo {author} {\bibfnamefont {R.~A.}\ \bibnamefont
  {Gallardo}}, \bibinfo {author} {\bibfnamefont {T.}~\bibnamefont {Schneider}},
  \bibinfo {author} {\bibfnamefont {S.}~\bibnamefont {Stienen}}, \bibinfo
  {author} {\bibfnamefont {A.}~\bibnamefont {Rold{\'{a}}n-Molina}}, \bibinfo
  {author} {\bibfnamefont {Y.}~\bibnamefont {Yuan}}, \bibinfo {author}
  {\bibfnamefont {K.}~\bibnamefont {Lenz}}, \bibinfo {author} {\bibfnamefont
  {J.}~\bibnamefont {Lindner}}, \bibinfo {author} {\bibfnamefont
  {P.}~\bibnamefont {Landeros}},\ and\ \bibinfo {author} {\bibfnamefont
  {J.}~\bibnamefont {Fassbender}},\ }\bibfield  {title} {\bibinfo {title}
  {Spin-wave modes in transition from a thin film to a full magnonic crystal},\
  }\href {https://doi.org/10.1103/physrevb.99.024426} {\bibfield  {journal}
  {\bibinfo  {journal} {Phys. Rev. B}\ }\textbf {\bibinfo {volume} {99}},\
  \bibinfo {pages} {024426} (\bibinfo {year} {2019})}\BibitemShut {NoStop}%
\bibitem [{\citenamefont {Lisenkov}\ \emph {et~al.}(2015)\citenamefont
  {Lisenkov}, \citenamefont {Kalyabin}, \citenamefont {Osokin}, \citenamefont
  {Klos}, \citenamefont {Krawczyk},\ and\ \citenamefont
  {Nikitov}}]{Lisenkov_2015}%
  \BibitemOpen
  \bibfield  {author} {\bibinfo {author} {\bibfnamefont {I.}~\bibnamefont
  {Lisenkov}}, \bibinfo {author} {\bibfnamefont {D.}~\bibnamefont {Kalyabin}},
  \bibinfo {author} {\bibfnamefont {S.}~\bibnamefont {Osokin}}, \bibinfo
  {author} {\bibfnamefont {J.~W.}\ \bibnamefont {Klos}}, \bibinfo {author}
  {\bibfnamefont {M.}~\bibnamefont {Krawczyk}},\ and\ \bibinfo {author}
  {\bibfnamefont {S.}~\bibnamefont {Nikitov}},\ }\bibfield  {title} {\bibinfo
  {title} {Nonreciprocity of edge modes in 1d magnonic crystal},\ }\href
  {https://doi.org/https://doi.org/10.1016/j.jmmm.2014.10.073} {\bibfield
  {journal} {\bibinfo  {journal} {J. Magn. Magn. Mater.}\ }\textbf {\bibinfo
  {volume} {378}},\ \bibinfo {pages} {313} (\bibinfo {year}
  {2015})}\BibitemShut {NoStop}%
\bibitem [{\citenamefont {Gallardo}\ \emph {et~al.}(2018)\citenamefont
  {Gallardo}, \citenamefont {Schneider}, \citenamefont {Rold\'an-Molina},
  \citenamefont {Langer}, \citenamefont {Fassbender}, \citenamefont {Lenz},
  \citenamefont {Lindner},\ and\ \citenamefont {Landeros}}]{Gallardo_2018}%
  \BibitemOpen
  \bibfield  {author} {\bibinfo {author} {\bibfnamefont {R.~A.}\ \bibnamefont
  {Gallardo}}, \bibinfo {author} {\bibfnamefont {T.}~\bibnamefont {Schneider}},
  \bibinfo {author} {\bibfnamefont {A.}~\bibnamefont {Rold\'an-Molina}},
  \bibinfo {author} {\bibfnamefont {M.}~\bibnamefont {Langer}}, \bibinfo
  {author} {\bibfnamefont {J.}~\bibnamefont {Fassbender}}, \bibinfo {author}
  {\bibfnamefont {K.}~\bibnamefont {Lenz}}, \bibinfo {author} {\bibfnamefont
  {J.}~\bibnamefont {Lindner}},\ and\ \bibinfo {author} {\bibfnamefont
  {P.}~\bibnamefont {Landeros}},\ }\bibfield  {title} {\bibinfo {title}
  {Dipolar interaction induced band gaps and flat modes in surface-modulated
  magnonic crystals},\ }\href
  {https://doi.org/https://doi.org/10.1103/PhysRevB.97.144405} {\bibfield
  {journal} {\bibinfo  {journal} {Phys. Rev. B}\ }\textbf {\bibinfo {volume}
  {97}},\ \bibinfo {pages} {144405} (\bibinfo {year} {2018})}\BibitemShut
  {NoStop}%
\bibitem [{\citenamefont {$\textstyle{\rm \mu MAG}$}()}]{SMMP}%
  \BibitemOpen
  \bibfield  {author} {\bibinfo {author} {\bibnamefont {$\textstyle{\rm \mu
  MAG}$}},\ }\href {https://www.ctcms.nist.gov/~rdm/mumag.org.html} {\bibinfo
  {journal} {{\color{black} Micromagnetic Modeling Activity Group,}
  www.ctcms.nist.gov/$\sim$rdm/mumag.org.html$\!\!$}\ }\BibitemShut {NoStop}%
\bibitem [{COM()}]{COMSOL}%
  \BibitemOpen
\bibfield  {journal} {  }\href {http://www.comsol.com} {}\bibinfo {note}
  {COMSOL Multiphysics$\textsuperscript{\textregistered}$~v.~5.4.~{\color{blue}
  www.comsol.com.} COMSOL AB, Stockholm, Sweden.}\BibitemShut {Stop}%
\bibitem [{\citenamefont {Mruczkiewicz}\ \emph {et~al.}(2013)\citenamefont
  {Mruczkiewicz}, \citenamefont {Krawczyk}, \citenamefont {Sakharov},
  \citenamefont {Khivintsev}, \citenamefont {Filimonov},\ and\ \citenamefont
  {Nikitov}}]{Mruczkiewicz_Comsol}%
  \BibitemOpen
  \bibfield  {author} {\bibinfo {author} {\bibfnamefont {M.}~\bibnamefont
  {Mruczkiewicz}}, \bibinfo {author} {\bibfnamefont {M.}~\bibnamefont
  {Krawczyk}}, \bibinfo {author} {\bibfnamefont {V.~K.}\ \bibnamefont
  {Sakharov}}, \bibinfo {author} {\bibfnamefont {Y.~V.}\ \bibnamefont
  {Khivintsev}}, \bibinfo {author} {\bibfnamefont {Y.~A.}\ \bibnamefont
  {Filimonov}},\ and\ \bibinfo {author} {\bibfnamefont {S.~A.}\ \bibnamefont
  {Nikitov}},\ }\bibfield  {title} {\bibinfo {title} {Standing spin waves in
  magnonic crystals},\ }\href {https://doi.org/10.1063/1.4793085} {\bibfield
  {journal} {\bibinfo  {journal} {J. Appl. Phys.}\ }\textbf {\bibinfo {volume}
  {113}},\ \bibinfo {pages} {093908} (\bibinfo {year} {2013})}\BibitemShut
  {NoStop}%
\bibitem [{\citenamefont {Rych{\l}y}\ \emph {et~al.}(2015)\citenamefont
  {Rych{\l}y}, \citenamefont {Gruszecki}, \citenamefont {Mruczkiewicz},
  \citenamefont {K{\l}os}, \citenamefont {Mamica},\ and\ \citenamefont
  {Krawczyk}}]{Rychly_2015}%
  \BibitemOpen
  \bibfield  {author} {\bibinfo {author} {\bibfnamefont {J.}~\bibnamefont
  {Rych{\l}y}}, \bibinfo {author} {\bibfnamefont {P.}~\bibnamefont
  {Gruszecki}}, \bibinfo {author} {\bibfnamefont {M.}~\bibnamefont
  {Mruczkiewicz}}, \bibinfo {author} {\bibfnamefont {J.~W.}\ \bibnamefont
  {K{\l}os}}, \bibinfo {author} {\bibfnamefont {S.}~\bibnamefont {Mamica}},\
  and\ \bibinfo {author} {\bibfnamefont {M.}~\bibnamefont {Krawczyk}},\
  }\bibfield  {title} {\bibinfo {title} {Magnonic crystals -- prospective
  structures for shaping spin waves in nanoscale},\ }\href
  {https://doi.org/10.1063/1.4932348} {\bibfield  {journal} {\bibinfo
  {journal} {Low Temperature Physics}\ }\textbf {\bibinfo {volume} {41}},\
  \bibinfo {pages} {745} (\bibinfo {year} {2015})}\BibitemShut {NoStop}%
\bibitem [{\citenamefont {van Kampen}\ \emph {et~al.}(2002)\citenamefont {van
  Kampen}, \citenamefont {Jozsa}, \citenamefont {Kohlhepp}, \citenamefont
  {LeClair}, \citenamefont {Lagae}, \citenamefont {de~Jonge},\ and\
  \citenamefont {Koopmans}}]{VanKampen_2002}%
  \BibitemOpen
  \bibfield  {author} {\bibinfo {author} {\bibfnamefont {M.}~\bibnamefont {van
  Kampen}}, \bibinfo {author} {\bibfnamefont {C.}~\bibnamefont {Jozsa}},
  \bibinfo {author} {\bibfnamefont {J.~T.}\ \bibnamefont {Kohlhepp}}, \bibinfo
  {author} {\bibfnamefont {P.}~\bibnamefont {LeClair}}, \bibinfo {author}
  {\bibfnamefont {L.}~\bibnamefont {Lagae}}, \bibinfo {author} {\bibfnamefont
  {W.~J.~M.}\ \bibnamefont {de~Jonge}},\ and\ \bibinfo {author} {\bibfnamefont
  {B.}~\bibnamefont {Koopmans}},\ }\bibfield  {title} {\bibinfo {title}
  {All-optical probe of coherent spin waves},\ }\href
  {https://doi.org/10.1103/PhysRevLett.88.227201} {\bibfield  {journal}
  {\bibinfo  {journal} {Phys. Rev. Lett.}\ }\textbf {\bibinfo {volume} {88}},\
  \bibinfo {pages} {227201} (\bibinfo {year} {2002})}\BibitemShut {NoStop}%
\bibitem [{\citenamefont {Kats}\ \emph {et~al.}(2016)\citenamefont {Kats},
  \citenamefont {Linnik}, \citenamefont {Salasyuk}, \citenamefont {Rushforth},
  \citenamefont {Wang}, \citenamefont {Wadley}, \citenamefont {Akimov},
  \citenamefont {Cavill}, \citenamefont {Holy}, \citenamefont {Kalashnikova},\
  and\ \citenamefont {Scherbakov}}]{Kats_2016}%
  \BibitemOpen
  \bibfield  {author} {\bibinfo {author} {\bibfnamefont {V.~N.}\ \bibnamefont
  {Kats}}, \bibinfo {author} {\bibfnamefont {T.~L.}\ \bibnamefont {Linnik}},
  \bibinfo {author} {\bibfnamefont {A.~S.}\ \bibnamefont {Salasyuk}}, \bibinfo
  {author} {\bibfnamefont {A.~W.}\ \bibnamefont {Rushforth}}, \bibinfo {author}
  {\bibfnamefont {M.}~\bibnamefont {Wang}}, \bibinfo {author} {\bibfnamefont
  {P.}~\bibnamefont {Wadley}}, \bibinfo {author} {\bibfnamefont {A.~V.}\
  \bibnamefont {Akimov}}, \bibinfo {author} {\bibfnamefont {S.~A.}\
  \bibnamefont {Cavill}}, \bibinfo {author} {\bibfnamefont {V.}~\bibnamefont
  {Holy}}, \bibinfo {author} {\bibfnamefont {A.~M.}\ \bibnamefont
  {Kalashnikova}},\ and\ \bibinfo {author} {\bibfnamefont {A.~V.}\ \bibnamefont
  {Scherbakov}},\ }\bibfield  {title} {\bibinfo {title} {Ultrafast changes of
  magnetic anisotropy driven by laser-generated coherent and noncoherent
  phonons in metallic films},\ }\href
  {https://doi.org/10.1103/PhysRevB.93.214422} {\bibfield  {journal} {\bibinfo
  {journal} {Phys. Rev. B}\ }\textbf {\bibinfo {volume} {93}},\ \bibinfo
  {pages} {214422} (\bibinfo {year} {2016})}\BibitemShut {NoStop}%
\bibitem [{\citenamefont {Salasyuk}\ \emph {et~al.}(2018)\citenamefont
  {Salasyuk}, \citenamefont {Rudkovskaya}, \citenamefont {Danilov},
  \citenamefont {Glavin}, \citenamefont {Kukhtaruk}, \citenamefont {Wang},
  \citenamefont {Rushforth}, \citenamefont {Nekludova}, \citenamefont
  {Sokolov}, \citenamefont {Elistratov}, \citenamefont {Yakovlev},
  \citenamefont {Bayer}, \citenamefont {Akimov},\ and\ \citenamefont
  {Scherbakov}}]{Salasyuk_2018}%
  \BibitemOpen
  \bibfield  {author} {\bibinfo {author} {\bibfnamefont {A.~S.}\ \bibnamefont
  {Salasyuk}}, \bibinfo {author} {\bibfnamefont {A.~V.}\ \bibnamefont
  {Rudkovskaya}}, \bibinfo {author} {\bibfnamefont {A.~P.}\ \bibnamefont
  {Danilov}}, \bibinfo {author} {\bibfnamefont {B.~A.}\ \bibnamefont {Glavin}},
  \bibinfo {author} {\bibfnamefont {S.~M.}\ \bibnamefont {Kukhtaruk}}, \bibinfo
  {author} {\bibfnamefont {M.}~\bibnamefont {Wang}}, \bibinfo {author}
  {\bibfnamefont {A.~W.}\ \bibnamefont {Rushforth}}, \bibinfo {author}
  {\bibfnamefont {P.~A.}\ \bibnamefont {Nekludova}}, \bibinfo {author}
  {\bibfnamefont {S.~V.}\ \bibnamefont {Sokolov}}, \bibinfo {author}
  {\bibfnamefont {A.~A.}\ \bibnamefont {Elistratov}}, \bibinfo {author}
  {\bibfnamefont {D.~R.}\ \bibnamefont {Yakovlev}}, \bibinfo {author}
  {\bibfnamefont {M.}~\bibnamefont {Bayer}}, \bibinfo {author} {\bibfnamefont
  {A.~V.}\ \bibnamefont {Akimov}},\ and\ \bibinfo {author} {\bibfnamefont
  {A.~V.}\ \bibnamefont {Scherbakov}},\ }\bibfield  {title} {\bibinfo {title}
  {Generation of a localized microwave magnetic field by coherent phonons in a
  ferromagnetic nanograting},\ }\href
  {https://doi.org/10.1103/physrevb.97.060404} {\bibfield  {journal} {\bibinfo
  {journal} {Phys. Rev. B}\ }\textbf {\bibinfo {volume} {97}},\ \bibinfo
  {pages} {060404(R)} (\bibinfo {year} {2018})}\BibitemShut {NoStop}%
\bibitem [{\citenamefont {Scherbakov}\ \emph {et~al.}(2019)\citenamefont
  {Scherbakov}, \citenamefont {Danilov}, \citenamefont {Godejohann},
  \citenamefont {Linnik}, \citenamefont {Glavin}, \citenamefont {Shelukhin},
  \citenamefont {Pattnaik}, \citenamefont {Wang}, \citenamefont {Rushforth},
  \citenamefont {Yakovlev}, \citenamefont {Akimov},\ and\ \citenamefont
  {Bayer}}]{Scherbakov_2019}%
  \BibitemOpen
  \bibfield  {author} {\bibinfo {author} {\bibfnamefont {A.~V.}\ \bibnamefont
  {Scherbakov}}, \bibinfo {author} {\bibfnamefont {A.~P.}\ \bibnamefont
  {Danilov}}, \bibinfo {author} {\bibfnamefont {F.}~\bibnamefont {Godejohann}},
  \bibinfo {author} {\bibfnamefont {T.~L.}\ \bibnamefont {Linnik}}, \bibinfo
  {author} {\bibfnamefont {B.~A.}\ \bibnamefont {Glavin}}, \bibinfo {author}
  {\bibfnamefont {L.~A.}\ \bibnamefont {Shelukhin}}, \bibinfo {author}
  {\bibfnamefont {D.~P.}\ \bibnamefont {Pattnaik}}, \bibinfo {author}
  {\bibfnamefont {M.}~\bibnamefont {Wang}}, \bibinfo {author} {\bibfnamefont
  {A.~W.}\ \bibnamefont {Rushforth}}, \bibinfo {author} {\bibfnamefont {D.~R.}\
  \bibnamefont {Yakovlev}}, \bibinfo {author} {\bibfnamefont {A.~V.}\
  \bibnamefont {Akimov}},\ and\ \bibinfo {author} {\bibfnamefont
  {M.}~\bibnamefont {Bayer}},\ }\bibfield  {title} {\bibinfo {title} {Optical
  excitation of single- and multimode magnetization precession in
  $\mathrm{Fe}$-$\mathrm{Ga}$ nanolayers},\ }\href
  {https://doi.org/10.1103/PhysRevApplied.11.031003} {\bibfield  {journal}
  {\bibinfo  {journal} {Phys. Rev. Applied}\ }\textbf {\bibinfo {volume}
  {11}},\ \bibinfo {pages} {031003} (\bibinfo {year} {2019})}\BibitemShut
  {NoStop}%
\end{thebibliography}%

\end{document}